\documentclass[a4paper,twocolumn,11pt,accepted=2022-10-13]{quantumarticle}
\pdfoutput=1

\usepackage{tikz}
\usepackage{fixltx2e}
\usepackage{dcolumn}
\usepackage{pgfplots}
\usepackage[arrowdel]{physics}
\usepackage[breaklinks=true,colorlinks=true,linkcolor=cyan,urlcolor=cyan,citecolor=blue]{hyperref}
\usetikzlibrary{graphs}
\usetikzlibrary{graphs.standard}
\usepackage{mathtools}
\usepackage{caption}
\usepackage{adjustbox}
\usepackage{amsfonts}
\usepackage{subcaption}
\usepackage{colortbl}
\usepackage{diagbox}
\usepackage[shortlabels]{enumitem}
\usepackage{qcircuit}
\usepackage[english]{babel}
\usepackage{algorithm}
\usepackage{dsfont}
\usepackage{color} 
\usepackage[framemethod=TikZ]{mdframed}
\usepackage{algorithm}
\usepackage{algpseudocode}
\usepackage{braket}
\usepackage{etoolbox}
\apptocmd{\sloppy}{\hbadness 10000\relax}{}{}
\usepackage[labeled]{multibib}
\usepackage{url}

\usepackage{tikz}
\usepackage{graphicx}
\usetikzlibrary{positioning}
\usepackage{wrapfig}
\usepackage{xfrac}
\usepackage{soul}
\usepackage{xcolor}
\DeclareUnicodeCharacter{2212}{-}
\makeatletter
\usepackage{cite}

\newcommand{\maxcut}{\mathsf{Max}\text{-}\mathsf{Cut}}

\newcommand{\secref}[1]{Section~\ref{#1}}

\newcommand{\tabref}[1]{Table~\ref{#1}}

\newcommand{\figref}[1]{Fig.~\ref{#1}}
\newcommand{\eqnref}[1]{Eq. (\ref{#1})}

\newcommand{\appref}[1]{Appendix~\ref{#1}}

\newcommand{\paramtheta}{\boldsymbol{\theta}}

\newcommand{\ansatz}{\textrm{ansatz}}

\newcommand{\QAOA}{\mathsf{QAOA}}

\newcommand{\LG}{\mathcal{L}(\mathcal{G})}
\newcommand{\G}{\mathcal{G}}

\newcommand{\argmin}{\text{arg}\text{min}}

\renewcommand{\>}{\rangle}
\newcommand{\<}{\langle}

\newcommand{\xbs}{\boldsymbol{x}}
\newcommand{\ybs}{\boldsymbol{y}}
\newcommand{\hbs}{\boldsymbol{h}}
\newcommand{\zbs}{\boldsymbol{z}}
\newcommand{\gbs}{\boldsymbol{g}}
\setstcolor{red}

\long\def\ca#1\cb{} 
\def\orcid#1{\kern -0.4em\href{https://orcid.org/#1}{\includegraphics[keepaspectratio,width=0.7em]{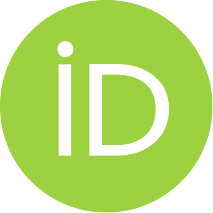}}}

\newcommand{\ZG}{\mathsf{Z}}

\newcommand{\XG}{\mathsf{X}}

\newcommand{\RX}{\mathsf{R}_x}

\newcommand{\RY}{\mathsf{R}_y}

\pgfplotsset{compat=1.17}

\begin{document}

\title{Graph neural network initialisation of quantum approximate optimisation}

\author{Nishant Jain~~\texorpdfstring{\orcid{0000-0003-1787-2481}}{}}
\affiliation{Indian Institute of Technology, Roorkee, India.}

\author{Brian Coyle~~\texorpdfstring{\orcid{0000-0002-3436-8458}}{}}
\affiliation{School of Informatics, University of Edinburgh, EH8 9AB Edinburgh, United Kingdom.}

\author{Elham Kashefi}
\affiliation{School of Informatics, University of Edinburgh, EH8 9AB Edinburgh, United Kingdom.}
\affiliation{LIP6, CNRS, Sorbonne Université, 4 place Jussieu, 75005 Paris, France.}

\author{Niraj Kumar~~\texorpdfstring{\orcid{0000-0003-3037-1083}}{}}
\affiliation{School of Informatics, University of Edinburgh, EH8 9AB Edinburgh, United Kingdom.}

\maketitle

\begin{abstract}
    Approximate combinatorial optimisation has emerged as one of the most promising application areas for quantum computers, particularly those in the near term. In this work, we focus on the quantum approximate optimisation algorithm ($\QAOA$) for solving the $\maxcut$ problem. Specifically, we address two problems in the $\QAOA$, how to initialise the algorithm, and how to subsequently train the parameters to find an optimal solution. For the former, we propose graph neural networks (GNNs) as a warm-starting technique for $\QAOA$. We demonstrate that merging GNNs with $\QAOA$ can outperform both approaches individually. Furthermore, we demonstrate how graph neural networks enables warm-start generalisation across not only graph instances, but also to increasing graph sizes, a feature not straightforwardly available to other warm-starting methods. For training the $\QAOA$, we test several optimisers for the $\maxcut$ problem up to $16$ qubits and benchmark against vanilla gradient descent. These include quantum aware/agnostic and machine learning based/neural optimisers. Examples of the latter include reinforcement and meta-learning. With the incorporation of these initialisation and optimisation toolkits, we demonstrate how the optimisation problems can be solved using $\QAOA$ in an end-to-end differentiable pipeline.
\end{abstract}

\section{Introduction}\label{sec:intro}
Among the forerunners for use cases of near-term quantum computers, dubbed \emph{noisy intermediate-scale} quantum (NISQ)~\cite{preskill_quantum_2018} are the \emph{variational} quantum algorithms (VQAs). The most well known of these is the variational quantum eigensolver~\cite{peruzzo_variational_2014} and the quantum approximate optimisation algorithm ($\QAOA$)~\cite{farhi_quantum_2014}. In their wake, many new algorithms have been proposed in this variational framework tackling problems in a variety of areas~\cite{mcclean_theory_2016, cerezo_variational_2021, bharti_noisy_2022}. The primary workhorse in such algorithms is typically the parameterised quantum circuit (PQC), and due to the heuristic and trainable nature of VQAs they have also become synonymous with `modern' quantum machine learning~\cite{mitarai_quantum_2018}. This is particularly evident with the adoption of PQCs as the quantum version of neural networks~\cite{farhi_classification_2018, benedetti_parameterized_2019}.

In this work, we focus on one particular VQA - the $\QAOA$ - primarily used for approximate discrete combinatorial optimisation. The canonical example of such a problem is finding the `maximum cut' ($\maxcut$) of a graph, where (for an unweighted graph) one aims to partition the graph nodes into two sets such that the sets have as many edges connecting them as possible. Discrete optimisation problems such as $\maxcut$ are extremely challenging to solve (specifically \textsf{NP}-\textsf{Hard}) and accurate solutions to such problems take exponential time in general. Aside from its theoretical relevance, $\maxcut$ finds applications across various fields such as study of the spin glass model, network design, VLSI and other circuit layout designs~\cite{barahona_application_1988}, and across data clustering~\cite{poland_clustering_2006}. While it is not believed quantum computers can solve \textsf{NP}-\textsf{Hard} problems \emph{efficiently}~\cite{nielsen_quantum_2010}, it is hoped that quantum algorithms such as $\QAOA$ may be able to outperform classical algorithms by some benchmark. Given the ubiquity of combinatorial optimisation problems in the real world, even incremental improvements may have large financial and quality impacts.

Due to this potential, there has been a rapid development in the study of the $\QAOA$ algorithm and its components, including (but not limited to) theoretical observations and limitations~\cite{hastings_classical_2019, farhi_quantum_2022, stilck_franca_limitations_2021, akshay_reachability_2020, boulebnane_improving_2020, akshay_parameter_2021, rabinovich_progress_2021, basso_quantum_2022}, variations on the circuit structure ($\ansatz$)~\cite{hadfield_quantum_2019, larose_mixer-phaser_2021, zhu_adaptive_2022, hadfield_analytical_2021, verdon_quantum_2019-1} used, the cost function~\cite{barkoutsos_improving_2020, kolotouros_evolving_2022, amaro_filtering_2022} and initialisation and optimisation methods~\cite{egger_warm-starting_2021, sack_quantum_2021, guerreschi_practical_2017, moll_quantum_2018, khairy_reinforcement-learning-based_2019, streif_training_2020} used for finding optimal solutions. Since the algorithm is suitable for near-term devices, there has also been substantial progress in experimental or numerical benchmarks~\cite{zhou_quantum_2020, streif_training_2020, amaro_case_2022, harrigan_quantum_2021, weidenfeller_scaling_2022} and the effect of quantum noise on the algorithm~\cite{xue_effects_2021, marshall_characterizing_2020}.

However, due to the limitations in running real experiments on small and unreliable NISQ devices, which currently are typically only accessible via expensive cloud computing platforms~\cite{larose_overview_2019}, it is important to limit the quantum resource (i.e. the overall number of runs, or the time for a single run on quantum hardware) required to solve a problem to the bare minimum. Therefore, effective initialisation and optimisation strategies for VQAs can dramatically accelerate the search for optimal problem solutions. The former ensures the algorithm begins `close' to a solution in the parameter space (near a local or global optimum), while the latter enables smooth and efficient traversal of the landscape. This is especially relevant given the existence of difficult optimisation landscapes in VQAs plagued by barren plateaus~\cite{mcclean_barren_2018, wiersema_exploring_2020, cerezo_cost_2021, larocca_diagnosing_2022}, local minima~\cite{you_exponentially_2021, rivera-dean_avoiding_2021} and narrow gorges~\cite{arrasmith_equivalence_2022}. To avoid these, and to enable efficient optimisation, several initialisation technique have been proposed for VQAs, including using tensor networks~\cite{dborin_matrix_2022}, meta-learning~\cite{verdon_learning_2019, sauvage_flip_2021, cervera-lierta_meta-variational_2021} and algorithm-specific techniques~\cite{egger_warm-starting_2021, sack_quantum_2021}.

Returning to the specifics of combinatorial optimisation, the use of machine and deep learning has been shown to be effective means of solving this family of problems, see for example Refs.~\cite{yao_experimental_2019,  cappart_combinatorial_2021,kotary_end--end_2021, schuetz_combinatorial_2022}. Of primary interest for our purposes are the works of~\cite{yao_experimental_2019, schuetz_combinatorial_2022}. The former~\cite{yao_experimental_2019} trains a \emph{graph neural network} (GNN) to solve $\maxcut$, while the latter~\cite{schuetz_combinatorial_2022} extends this to more general optimisation problems, and demonstrates scaling up to millions of variables. Based on these insights and the recent trend in the quantum domain of incorporating VQAs with neural networks (with software libraries developed for this purpose~\cite{bergholm_pennylane_2022, broughton_tensorflow_2020}) indicates that using both classical and quantum learning architectures synergistically has much promise. We extend this hybridisation in this work.

This paper is divided into two parts. In the first part (Sections~\ref{sec:warm_start_qaoa}-\ref{sec:initialisation_numerics}), we discuss the previous works in $\QAOA$ initialisation and give our first contribution: an initialisation strategy using graph neural networks. Specifically, we merge GNN solvers with the warm-starting technique for $\QAOA$ of~\cite{egger_warm-starting_2021}, and demonstrate the effectiveness of this via numerical results in \secref{sec:initialisation_numerics}. By then examining the conclusion of~\cite{schuetz_combinatorial_2022}, we can see how our GNN approach would allow $\QAOA$ initialisation to scale far beyond the capabilities of current generation near-term quantum devices. In the second part of the paper (\secref{sec:qaoa_optimisation}), we then complement this by evaluating several methods of optimisation techniques for the $\QAOA$ proposed in the literature, including quantum-aware, quantum-agnostic and neural network based optimisation approaches.

\begin{figure*}[!htb]
    \centering
    \includegraphics[width=\textwidth, height=0.4\textwidth]{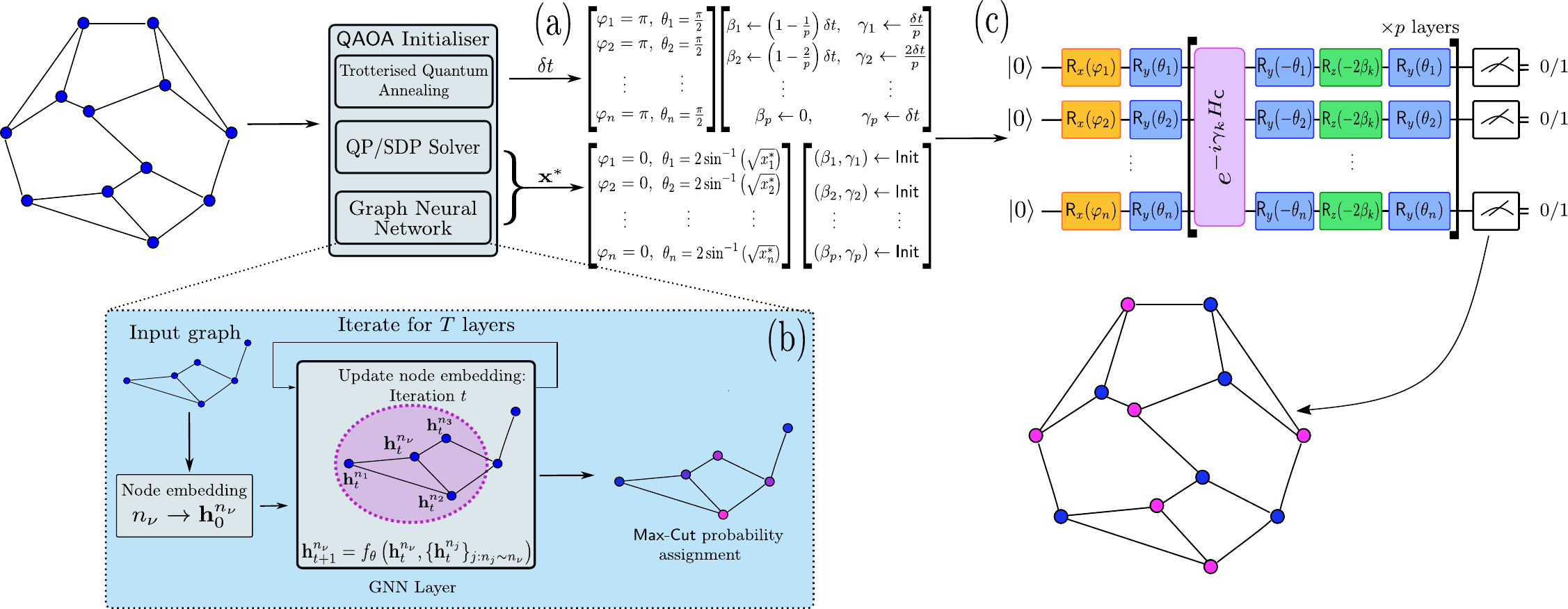}
    \caption{\textbf{Initialisation techniques for $\QAOA$.} (a) This part highlights the difference between the warm-starting techniques (using semi-definite programming (SDP) relaxations~\cite{egger_warm-starting_2021} or graph neural networks (GNNs)) and Trotterised quantum annealing~\cite{sack_quantum_2021} (TQA). TQA produces initial angles, $\{\boldsymbol{\beta}, \boldsymbol{\gamma}\}$, whereas warm-starting techniques initialise the $\QAOA$ state. `$\textsf{Init}$' here refers to any parameter initialisation scheme. In this work, we choose a specific initialisation technique called the Xavier initialisation \cite{glorot_understanding_2010} for warm-starting techniques. (b) GNNs take an embedding of the initial graph and applies updates to the embeddings based on the neighbours of each node, using an parameterised function, $f_{\paramtheta}$. In this case, the GNN outputs a probability for each node being on either side of the cut. (c) A unified $p$-layer $\QAOA$ circuit for all initialisation schemes. In TQA, fixed choices for angles $\theta, \phi$ initialise vanilla $\QAOA$ with the standard mixer, whereas warm-starting produces an initial state and mixer encoding a probabilistic solution given by the SDP or GNN. Here, $\boldsymbol{x}^*$ implicitly encodes the regularisation parameter $\epsilon$ discussed in~\cite{egger_warm-starting_2021}. Note that both the GNN and $\QAOA$ parameters can be trained in an end-to-end differentiable manner, in contrast to other schemes.}
    \label{fig:init_methods_overview}
\end{figure*}

\subsection{\texorpdfstring{$\QAOA$}{} for solving \texorpdfstring{$\maxcut$}{}}

For concreteness in this work, we focus on the discrete optimisation problem known as $\maxcut$. It involves finding a division (a cut) of the vertices of a (weighted) graph into two sets, which maximises the sum of the weights over all the edges across the vertex subsets. For unweighted graphs, this cut will maximise simply the number of edges across the two subsets.

The problem can be recast to minimising the weighted sum of operators acting on the vertices of a given graph. Mathematically, this can be stated as follows. Given a graph $\mathcal{G} := \left(\mathcal{V}, \mathcal{E}\right)$ with vertices $\mathcal{V}$ and edges $\mathcal{E} = \{(i, j) | i, j \in \mathcal{V} \text{ and } i \neq j\}$, the $\maxcut$ can be found by minimising the following cost function,

\begin{equation} \label{eqn:maxcut_cost_function_classical}
    \mathsf{C}(\boldsymbol{z}) = -\sum_{\langle i,j\rangle\in \mathcal{E}}w_{ij}(1-z_{i}z_{j})
\end{equation}

where $\boldsymbol{z} := z_1z_2\dots z_n$ are variables for each vertex, $i$, such that $z_i \in \{+1, -1\}$ and $w_{ij}$ is the corresponding weight of the edge between vertices $i$ and $j$. In this case, the value (sign) of $z_i$ determines on which side of the cut the node resides. The $\maxcut$ problem is a canonical \textsf{NP}-complete problem~\cite{garey_computers_1990}, meaning there is no known efficient polynomial time algorithm in general. Further, $\maxcut$ is also known to be \textsf{APX}-hard \cite{papadimitriou_optimization_1991}, meaning there is also no known polynomial time \emph{approximate} algorithm. The current best known polynomial time approximate classical approach is the Goemans-Williamson (GW) algorithm which is able to achieve an \emph{approximation ratio}, $r \approx 0.878$, where:
\begin{align} \label{eqn:definition_approx_ratio}
    r &= \frac{\text{Approximate cut}}{\text{Optimal cut}}\\
    &= \frac{2}{\pi}\left(\underset{0 \le \theta \le \pi}{\text{min}}\frac{\theta}{1 -\cos\theta}\right) \approx 0.878
\end{align}
Assuming the unique games conjecture (UGC)~\cite{khot_power_2002, khot_optimal_2007} classically, then the GW algorithm achieves the best approximation ratio for $\maxcut$. Without this conjecture, it has been proven that it is \textsf{NP}-hard to approximate the $\maxcut$ value with an approximation ratio better than $\smash{\frac{16}{17} \approx 0.941}$.

To address $\maxcut$ quantum mechanically, one can quantise the cost function~\eqnref{eqn:maxcut_cost_function_classical} by replacing the variables with \emph{operators}, $z_i \rightarrow \ZG_i$, where $\ZG$ is the Pauli-$\ZG$ matrix. The cost function can now be described with a \emph{Hamiltonian}:
\begin{equation} \label{eqn:maxcut_hamiltonian_quantum}
    H_{\mathsf{C}}(\boldsymbol{z}) = \sum_{\langle i,j\rangle\in \mathcal{E}}w_{ij}(1-\ZG_{i}\ZG_{j})
\end{equation}
where the actual cost - corresponding to the cut size - is extracted as the expectation value of this Hamiltonian with respect to a quantum state, $\ket{\psi}$:
\begin{equation} \label{eqn:maxcut_cost_function_quantum}
    \mathsf{C}(\boldsymbol{z}) := -\bra{\psi}H_{\mathsf{C}}\ket{\psi}
\end{equation}
The goal of a quantum algorithm is then to find the \emph{ground} state, $\ket{\psi}_G := \argmin_{\ket{\psi}} \bra{\psi}H_{\mathsf{C}}\ket{\psi}$, i.e. the state which minimises the \emph{energy} of the Hamiltonian, $H_{\mathsf{C}}$. Constructing the Hamiltonian as in~\eqnref{eqn:maxcut_hamiltonian_quantum}, ensures that the minimum energy state is exactly the state encoding the $\maxcut$ of the problem graph:
\begin{equation}
    \ket{\psi}_G = \ket{\psi}_{\maxcut}
\end{equation}

However, since the $\maxcut$ problem is \textsf{NP}-\textsf{Hard}, we expect that finding this ground state, $\ket{\psi}_{\maxcut}$, will also be hard in general. The $\QAOA$ algorithm attempts to solve this by initialising with respect to an \emph{easy} Hamiltonian (also called a `mixer' Hamiltonian):
\begin{equation} \label{eqn:qaoa_mixer_hamiltonian}
    H_{\mathsf{M}} = \sum_{i=1}^n\XG_{i}
\end{equation}

which has as an eigenstate the simple product state, $\ket{\psi}_{\text{init}} = \ket{+}^{\otimes n} = \mathsf{H}\ket{0}^{\otimes n}$ where $\mathsf{X}$ and $\mathsf{H}$ are the Pauli-$\mathsf{X}$ and Hadamard operators respectively. This can be viewed as an initialisation which is a superposition of all possible candidate solutions. The $\QAOA$ then attempts to simulate adiabatic evolution from $\ket{\psi}_{\text{init}}$ to the target state $\ket{\psi}_{\maxcut}$ by an alternating bang-bang application of two unitaries derived from the Hamiltonians,~\eqnref{eqn:maxcut_hamiltonian_quantum}, \eqnref{eqn:qaoa_mixer_hamiltonian}, which are respectively:
\begin{equation} \label{eqn:qaoa_cost_mixer_unitaries}
    U_{\mathsf{C}}(\gamma) = e^{-i\gamma H_{\mathsf{C}}} ~\text{ and }~     U_{\mathsf{M}}(\beta) = e^{-i\beta H_{\mathsf{M}}}
\end{equation}

In the $\QAOA$, the parameters, $\gamma, \beta$, are trainable, and govern the length of time each operator is applied for. These two unitaries are alternated in $p$ `layers' acting on the initial state, so the final state is prepared using $2p$ parameters, $\{\boldsymbol{\beta},\boldsymbol{\gamma}\} := \{\beta_1, \beta_2, \dots, \beta_p, \gamma_1, \gamma_2, \dots, \gamma_p\}$:
\begin{equation} \label{eqn:vanilla_qaoa_final_state}
  \ket{\psi_{\boldsymbol{\beta},\boldsymbol{\gamma}}} = U_{\mathsf{M}}(\beta_p)U_{\mathsf{C}}(\gamma_p)\dots U_{\mathsf{M}}(\beta_{1})U_{\mathsf{C}}(\gamma_{1})\ket{+}^{\otimes n}
\end{equation}

Optimising the parameters, $\{\boldsymbol{\beta},\boldsymbol{\gamma}\}$ serves as a proxy for finding the ground state, and so we aim that after a finite depth, $p$, we achieve a state $\ket{\psi_{\boldsymbol{\beta},\boldsymbol{\gamma}}}$ which is close to the target state $\ket{\psi}_{\maxcut}$.

\section{Initialising the \texorpdfstring{$\QAOA$}{}} \label{sec:warm_start_qaoa}

Since searching over the non-convex parameter landscape for an optimal setting of the $\{\boldsymbol{\gamma}, \boldsymbol{\beta}\}$ parameters directly on quantum hardware may be expensive and/or challenging, any attempts to initialise the $\QAOA$ parameters near a candidate solution are extremely valuable as the algorithm would then begin its search from an already good approximate solution. Such approaches are dubbed as `\emph{warm-starts}'~\cite{egger_warm-starting_2021}, in contrast to `cold-starts'. One could consider a cold-start to be a random initialisation of $\{\boldsymbol{\gamma}, \boldsymbol{\beta}\}$, or by using an initial state which encodes no problem information, e.g. $\ket{+}^{\otimes n}$, as in vanilla $\QAOA$. In this work, we refer to cold-start as the latter, and `random initialisation' to mean a random setting of the parameters, $\{\boldsymbol{\gamma}, \boldsymbol{\beta}\}$. We first revisit and summarise two previous approaches~\cite{egger_warm-starting_2021, sack_quantum_2021}, before presenting our approach to $\QAOA$ initialisation. We illustrate these previous initialisation approaches in \figref{fig:init_methods_overview}, which we review briefly in~\secref{sec:warm_start_qaoa_egger} and~\secref{sec:warm_start_qaoa_sack}, and also our approach based on graph neural networks, which we introduce in~\secref{sec:warm_start_qaoa_GNN}. For simplicity, we focus on the simplest version of $\QAOA$, but the methods could be extended to other variants, for example recursive $\QAOA$ (\textsf{RQAOA})~\cite{egger_warm-starting_2021, bravyi_hybrid_2022, bravyi_obstacles_2020}.

\subsection{Continuous relaxations} \label{sec:warm_start_qaoa_egger}

The first approach of~\cite{egger_warm-starting_2021} proposed a warm-start for $\QAOA$ method which can be applied to $\maxcut$ as a special case of a quadratic unconstrained binary optimisation (\textsf{QUBO}) problem. In this work, two sub-approaches were discussed. The first converts the \textsf{QUBO} into its continuous quadratic relaxation form which is efficiently solvable and directly uses the output of this relaxed problem to initialise the $\QAOA$ circuit. The second approach applies the random-hyperplane rounding method of the GW algorithm to generate a candidate solution for the \textsf{QUBO}. 
For $\maxcut$, this \textsf{QUBO} can be written in terms of the graph Laplacian, $\mathcal{L}_{\mathcal{G}} = \mathcal{D} - \mathcal{A}$ (where $\mathcal{D}$ is the diagonal degree matrix, and $\mathcal{A}$ is the adjacency matrix of $\mathcal{G}$) as follows (we utilise this form later in this work):
\begin{equation}\label{eqn:qubo_formalism_maxcut_laplacian}
     \max_{\zbs \in \{-1, 1\}^n} \zbs^T \mathcal{L}_{\mathcal{G}} \zbs
\end{equation}
However, by removing the requirement that each $\zbs_i$ is binary, one can obtain an efficiently solvable continuous relaxation that can serve as warm-start for solving~\eqnref{eqn:qubo_formalism_maxcut_laplacian}. Since $\mathcal{L}_\mathcal{G}$ is a positive semidefinite (PSD) matrix, the relaxed form can be trivially written as,
\begin{equation}\label{eqn:continuous_qubo_formalism}
    \max_{\xbs \in [0, 1]^n} (2\xbs-1)^T \mathcal{L}_\mathcal{G} (2\xbs-1),
\end{equation}
using the translation $\zbs_i \in \{-1, 1\} \rightarrow \zbs_i = (2\xbs_i-1),~\xbs_i \in \{0, 1\}$ and then allowing $\xbs_i$ to be continuous in this interval. If the matrix in the \textsf{QUBO} is not PSD however, then one can obtain another continuous relaxation, as a semidefinite programme (SDP)~\cite{overton_semidefinite_1997, egger_warm-starting_2021}. The output of this optimisation is a real vector $\xbs^*$ which, when a rounding procedure is performed (i.e. the GW algorithm), is a candidate solution for the original $\maxcut$. In order to use this relaxed solution to initialise $\QAOA$, Ref.~\cite{egger_warm-starting_2021} also demonstrated that the initial state from~\eqnref{eqn:vanilla_qaoa_final_state} and the mixer Hamiltonian~\eqnref{eqn:qaoa_mixer_hamiltonian} must also be altered as:
\begin{equation} \label{eqn:warm_start_initial_mixer}
    \begin{split}
        &\ket{\psi}_{\text{init}}^{\text{CS}} = \ket{+}^{\otimes n} \rightarrow \ket{\psi}_{\text{init}}^{\text{WS}} = \bigotimes\limits_{i=1}^n \RY(\paramtheta_i) \ket{0}^{\otimes n}.\\
          &H_{\mathsf{M}} \rightarrow \sum\limits_{i=1}^n H^{\text{WS}}_{\mathsf{M}, i},\\
          &H^{\text{WS}}_{\mathsf{M}, i} = \left(
          \begin{array}{cc}
             2\xbs^*_i - 1  & -2\sqrt{\xbs_i^*(1-\xbs_i^*)} \\
              -2\sqrt{\xbs_i^*(1 - \xbs_i^*)}  & 1 - 2\xbs^*_i
          \end{array}
          \right)
    \end{split}
\end{equation}
where $\paramtheta_i = 2\sin^{-1}(\sqrt{\xbs^*_i})$. One can immediately see that $\ket{\psi}_{\text{init}}^{\text{WS}}$ is the ground state of $H_{\mathsf{M}}$ with eigenvalue $-n$. One possible issue that may arise with this warm-start is if the relaxed solution $\xbs_i^*$ is either 0 or 1. When this happens, the qubit $i$ would be initialised to state $\ket{0}$ or $\ket{1}$, respectively. This means the qubit would be unaffected by the problem Hamiltonian~\eqnref{eqn:maxcut_hamiltonian_quantum} which only contains Pauli $\textsf{Z}$ terms. To account for this possibility, ~\cite{egger_warm-starting_2021} modifies $\paramtheta_i$ in~\eqnref{eqn:warm_start_initial_mixer} with a regularisation parameter, $\epsilon \in [0, 0.5]$ if the candidate solution, $\xbs^*_i$ is too close to $0$ or $1$.

Examining~\figref{fig:init_methods_overview}, this initialisation scheme is achieved by setting the angles in the initial state, $\boldsymbol{\varphi}_i = 0$ and $\paramtheta_i = 2\sin^{-1}(\sqrt{\xbs^*_i})~\forall i$ where the initial state can be expressed as $ \ket{\psi}_{\text{init}}^{\text{WS}} = \bigotimes \RX(\boldsymbol{\varphi}_i)\RY(\paramtheta_i) \ket{0}^{\otimes n} $. The `\textsf{Init}' in this figure implies that one is free to choose any $\QAOA$ parameter initialisation method as this warm-start approach only modifies the input state and the mixer Hamiltonian.

\subsection{Trotterised quantum annealing} \label{sec:warm_start_qaoa_sack}
A second proposed method~\cite{sack_quantum_2021} to initialise $\QAOA$ uses concepts from quantum annealing~\cite{kadowaki_quantum_1998, hauke_perspectives_2020}, which is a popular method of solving \textsf{QUBO} problems of the form~\eqnref{eqn:qubo_formalism_maxcut_laplacian}. $\QAOA$ was proposed as a discrete gate-based method to emulate quantum adiabatic evolution, or quantum annealing. Therefore, one may hope that insights from quantum annealing may be useful in setting the initial angles for the $\QAOA$ circuit parameters. In a method proposed by~\cite{sack_quantum_2021} (dubbed Trotterised quantum annealing (TQA)), one fixes the $\QAOA$ circuit depth, $p$, and sets the parameters as:
\begin{equation} \label{eqn:tqa_initialisation}
    \boldsymbol{\gamma}_k = \frac{k}{p}\delta t \qquad        \boldsymbol{\beta}_k = \left(1-\frac{k}{p}\right)\delta t
\end{equation}
where $k = 1, \cdots, p$ and $\delta t$ is a time interval which is a-priori unknown and given as a fraction of the (unknown) total optimal anneal time, $T^*$, resulting in $\delta t= T^*/p$. The authors of~\cite{sack_quantum_2021} observed an optimal time step value $\delta t$ to be $\approx 0.75$ for 3-regular graphs and $\mathcal{O}(1)$ for other graph ensembles. In~\cite{sack_quantum_2021}, the $\QAOA$ was initialised with $\delta t = 0.75$ and observed to help avoid local minima and find near-optimum minima close to the global minimum. We also choose this value generically in our numerical results later in the text, although one should ideally pre-optimise the value for each graph instance.

Note that in contrast to the warm-starting method from the previous section, the TQA approach initialises the \emph{parameters}, $\{\boldsymbol{\beta}, \boldsymbol{\gamma}\}$ rather than the initial $\QAOA$ state (and mixer Hamiltonian) which is set as in vanilla $\QAOA$ as $\ket{+}$. Again, revisiting~\figref{fig:init_methods_overview}, this initial state can be achieved by choosing $\boldsymbol{\varphi}_j = \pi, \paramtheta_j = \sfrac{\pi}{2},~\forall j$. This is due to the fact that $\mathsf{R}_y\left(\sfrac{\pi}{2}\right)\mathsf{R}_x(\pi)\ket{0} \propto \mathsf{X}\mathsf{R}_x\left(\sfrac{\pi}{2}\right)\ket{0} = \mathsf{H}\ket{0} = \ket{+}$. Similarly, for the mixer Hamiltonian, we have $\mathsf{R}_y\left(\sfrac{\pi}{2}\right)\mathsf{R}_z(-2\beta_k)\mathsf{R}_y\left(-\sfrac{\pi}{2}\right) \propto \mathsf{H}\mathsf{X}\mathsf{R}_z(-2\beta_k)\mathsf{X}\mathsf{H} = \mathsf{H}\mathsf{R}_z(2\beta_k)\mathsf{H} = \mathsf{R}_x(2\beta_k)$, which is the single qubit mixer unitary from vanilla $\QAOA$, up to a redefinition of $\beta_k$.

\section{Graph neural network warm-starting of \texorpdfstring{$\QAOA$}{}} \label{sec:warm_start_qaoa_GNN}

Now that we have introduced the $\QAOA$, and alternative methods for warm-starting its initial state and/or initial algorithm parameters, let us turn now to our proposed method; the use of \emph{graph neural networks}. This approach is closest to the relaxation method of~\secref{sec:warm_start_qaoa_egger} in that the GNN provides an alternative initial state to vanilla $\QAOA$, and so it is to this method which we primarily compare. One of the main drawbacks of using SDP relaxations and the GW algorithm, is that \emph{every} graph for which the $\maxcut$ must be found generates a new problem instance to be initialised. However, on the other hand, the GW algorithm comes equipped with performance guarantees (generating an approximate solution within $88\%$ of the optimal answer).

As we shall see, using graph neural networks as an initialiser allows a generalisation across many graph instances at once. Importantly, even increasing the number of qubits will not significantly affect the time complexity of such approaches as it can be interpreted as a learned prior over graphs. We also demonstrate how the model can be trained on a small number of qubits, and still perform well on larger problem instances (size generalisation), a feature not present in any of the previous initialisation methods for $\QAOA$. Furthermore, the incorporation of a differentiable initialisation structure allows the \emph{entire} pipeline of $\QAOA$ to become \emph{end-to-end} differentiable, which is particularly advantageous since it makes the problem ameanable to the automatic differentiation functionality of many deep learning libraries~\cite{paszke_pytorch_2019, abadi_tensorflow_2016}. First, let us begin by introducing graph neural networks.

\subsection{Graph neural networks}\label{ssec:graph_nns_background}

Graph neural networks (GNNs)~\cite{scarselli_graph_2009} are a specific neural network model designed to operate on graph-structured data, and typically function via some message passing process across the graph. They are example models in \emph{geometric} deep learning~\cite{bronstein_geometric_2021}, where one incorporates problem symmetries and invariants into the learning protocol. Examples have also been proposed in the field of quantum machine learning~\cite{verdon_learning_2019, verdon_quantum_2019-2, larocca_group-invariant_2022, skolik_equivariant_2022}. Specifically, graph neural networks operate by taking an encoding of the input graph describing a problem of interest and outputting a transformed encoding. Each graph node is initially encoded as a vector, which are then updated by the GNN to incorporate information about the relative features of the graph in the node. This is done by taking into account the connections and directions of the edges within the graph, using quantities such as the node degree, and the adjacency matrix. This transformed graph information is then used to solve the problem of interest. There are many possible architectures for how this graph information is utilised in the GNN, including attention based mechanisms~\cite{velickovic_graph_2018} or graph convolutions~\cite{zhang_graph_2019} for example, see e.g. Ref.~\cite{zhou_graph_2020} for a review.

In order to transform the feature embeddings, the GNN is trained for a certain number of iterations (a hyperparameter). For a given graph node, $n_\nu$, we associate a vector, $\hbs^{n_\nu}_t$, where $t$ is the current iteration. In the next iteration ($t+1$), to update the vector for node $n_\nu$, we first compute some function of the vector embeddings of the nodes in a neighbourhood of $n_\nu$, denoted as $\mathcal{N}(n_\nu) = \{n_j\}_{j: n_j \sim n_\nu}$. A-priori, there is no limitation on how large this neighbourhood can be, making it larger will increase training time and difficulty, but increase representational power. These function values are then aggregated (for example by taking an average) and combined with the node vector (with perhaps a non-linear activation function) at the previous iteration to generate $\hbs^{n_{\nu}}_{t+1}$. Each nodes update increases the information contained relative to a larger subset of nodes in the graph. The collective action of these operations can be described by a parameterised, trainable function, $f_{\paramtheta}(\hbs^{n_\nu}_t, \{\hbs^{n_{j}}_t\}_{j: n_j \sim n_\nu})$ (see~\figref{fig:init_methods_overview}), whose parameters, $\paramtheta$, are suitably trained to minimise a cost function. In all cases here, we initialise all the elements of the feature vectors, $\hbs^{n_\nu}_0$ to be the degree of the node, $n_{\nu}$. For the specific GNN architecture we use in the majority of this work, we choose the \emph{line} graph neural network (LGNN)~\cite{chen_supervised_2019}, shown to be competitive on combinatorial optimisation problems~\cite{yao_experimental_2019}. However we also incorporate the graph convolutional network (GCN) proposed for combinatorial optimisation by~\cite{schuetz_combinatorial_2022} in some numerical results in~\secref{sec:initialisation_numerics}. We give further details about these two architectures in~\appref{app:gnn_architectures}. 

Once we have a trained GNN for a certain number of iterations, $T$, we can use the information encoded in $\{\hbs^{n_j}_{T}\}_j$ for the problem at hand. A simple example would be to attach a multi-layer perception and perform classification on each node, where $\{\hbs^{n_j}_{T}\}$ behaves as feature vectors encoding the graph structure. For our purposes, we use these vectors to generate probabilities on the nodes. These are the probabilities that the node is in a given side of the $\maxcut$, which are then taken as the values $\xbs^*$ in warm-started $\QAOA$ circuit and its mixer Hamiltonian~\eqnref{eqn:warm_start_initial_mixer}.

\subsection{Graph neural networks for \texorpdfstring{$\maxcut$}{}}

To attach this probability, there are at least two possible methods one could apply. Firstly, one could consider using reinforcement learning or long short term memories (LSTMs)~\cite{khalil_learning_2017, deudon_learning_2018}. These methods generate probabilities in a step-wise fashion by employing a sequential dependency. To train these, one may use a policy gradient method~\cite{kool_attention_2019} (dubbed as an `\emph{autoregressive decoding}' approach~\cite{joshi_learning_2021}).

The second (simpler) method is to treat edge independently, and generate the probability of each edge being present in the $\maxcut$ or not (a `non-autoregressive decoding'). This can be formulated as a vector, $\boldsymbol{p}$, where each element corresponds to a node, $n_{\nu}$, generated by applying a softmax to the final output feature vectors of the GNN, $\{\hbs^{n_j}_{T}\}_j$~\cite{yao_experimental_2019, schuetz_combinatorial_2022}:
\begin{equation} \label{eqn:gnn_maxcut_probs}
    p_{n_{\nu}}(\boldsymbol{\theta}) = \frac{\exp\left(\hbs^{n_{\nu}, 0}_{T}\right)}{\sum_{j \in \{0, 1\}} \exp\left(\hbs^{n_{\nu}, j}_{T}\right)}
\end{equation}
In~\cite{yao_experimental_2019}, for each node, $n_{\nu}$, the final output is the two dimensional vector $[\hbs^{n_{\nu}, 0}_{T}, \hbs^{n_{\nu}, 1}_{T}]$, constructed as the output of a final two-output linear layer. The probability for each node is then taken as one of these outputs (say $j=0$) via the softmax in~\eqnref{eqn:gnn_maxcut_probs} and then used in the cost function described in the next section.

\subsubsection{Unsupervised training} \label{ssec:unsupervised_training_gnn}

Now that we have defined the structure and output of the GNN, it must be suitably trained. One approach is to use supervised training, however this may require a large number of example graphs to serve as the ground truth. Instead, following~\cite{yao_experimental_2019, schuetz_combinatorial_2022}, we opt for an \emph{unsupervised} approach, bypassing the need for labels. To do so, we choose the cost function as~\cite{yao_experimental_2019}, which is given by the $\maxcut$ \textsf{QUBO} itself in terms of the graph Laplacian (\eqnref{eqn:qubo_formalism_maxcut_laplacian}):
\begin{align}\label{eqn:gnn_cost_function}
     \mathsf{C}_{\text{GNN}} = -\min_{\paramtheta}\frac{1}{T} \sum\limits_{t=1}^T (-\mathsf{C}^t_{\text{GNN}}(\paramtheta)) \\
     \mathsf{C}^t_{\text{GNN}}(\paramtheta) = \frac{1}{4}(2\boldsymbol{p}-1)^T\mathcal{L}_{\mathcal{G}}(2\boldsymbol{p}-1)
\end{align}
where $\boldsymbol{p}\in [0, 1]^n$ is the probability vector from the GNN~\eqnref{eqn:gnn_maxcut_probs}. In~\eqnref{eqn:gnn_cost_function}, we define the cost function as an average over a training set of $T$ graphs, $\{\mathcal{G}_{t}\}_{t=1}^T$. Note, that the graphs in the training set do not have to be the same size as the graphs of interest; the GNN can be trained on an ensemble of graphs of different sizes, or graphs which are strictly smaller (or larger) than the test graph. We utilise this feature to improve GNN training in~\secref{ssec:gnn_versus_gw_numerical} and to demonstrate the generalisation capabilities of the GNN in~\secref{ssec:gnn_generalisation}. See Ref.~\cite{schuetz_combinatorial_2022} for how the cost function and GNN structure could be adapted to alternate \textsf{QUBO} type problems.

\section{Initialisation numerical results} \label{sec:initialisation_numerics}

Let us first study the impact of the initialisation schemes discussed above on the $\QAOA$ numerically. In all of the below, the approximation ratio, $r$, will be the figure of merit. We also use Xavier initialisation~\cite{glorot_understanding_2010} for the $\QAOA$ parameters in all cases except for the TQA initialisation method. This initialises each parameter from a uniform distribution over an interval depending on the number of parameters in the $\QAOA$ circuit.

\begin{figure}[!ht]
\begin{tikzpicture}
\begin{axis}[
    xlabel={Cut size (\% of true $\maxcut$)},
    ylabel={\% of graphs},
    ylabel near ticks,
    legend cell align={left},
    ymin=0, ymax=25,
    minor y tick num = 3,
    area style,
    legend pos=north west,
    ymajorgrids=true,
    grid =major,
    height = 6cm,
    width  = 7cm
    ]
\addplot+[ybar interval,mark=no] plot coordinates { (80,10) (85, 15) (90, 10) (95, 10) (100, 5)  };
\addplot+[ybar interval,mark=no] plot coordinates { (70,15) (75, 5) (80, 10) (85, 10) (90, 5) (95, 5)  };

\legend{LGNN $\QAOA$, Cold-start $\QAOA$}
\end{axis}
\end{tikzpicture}
\caption{\textbf{Success probability of the graph neural network on $3$-regular graphs for $\maxcut$.}
Histogram shows the number of graphs on which GNN $\QAOA$ versus cold-start $\QAOA$ can achieve a certain ratio of the optimal cut. Here, we set $p=5$ and $n=12$ qubits and generate the percentages over $50$ random graphs. The GNN initialised version is able to generate larger cut values than the vanilla version of $\QAOA$.}
\label{fig:histogram_gnn_v_coldstart}
\end{figure}
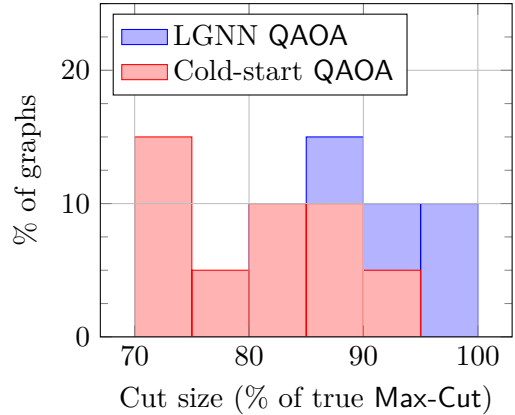

We begin by benchmarking the graph neural network $\QAOA$ against a random cold start initialisation for the two architectures discussed above, the line graph neural network (LGNN) and the graph convolutional network (GCN). To demonstrate feasibility, \figref{fig:histogram_gnn_v_coldstart} shows the success probability of the vanilla (cold-started) $\QAOA$ against the (LGNN-)$\QAOA$ initialisation over $50$ random $3$-regular graphs with $12$ qubits and $p=5$ $\QAOA$ depth. We observe the GNN-$\QAOA$ is capable of generating larger cuts on average than the vanilla version.

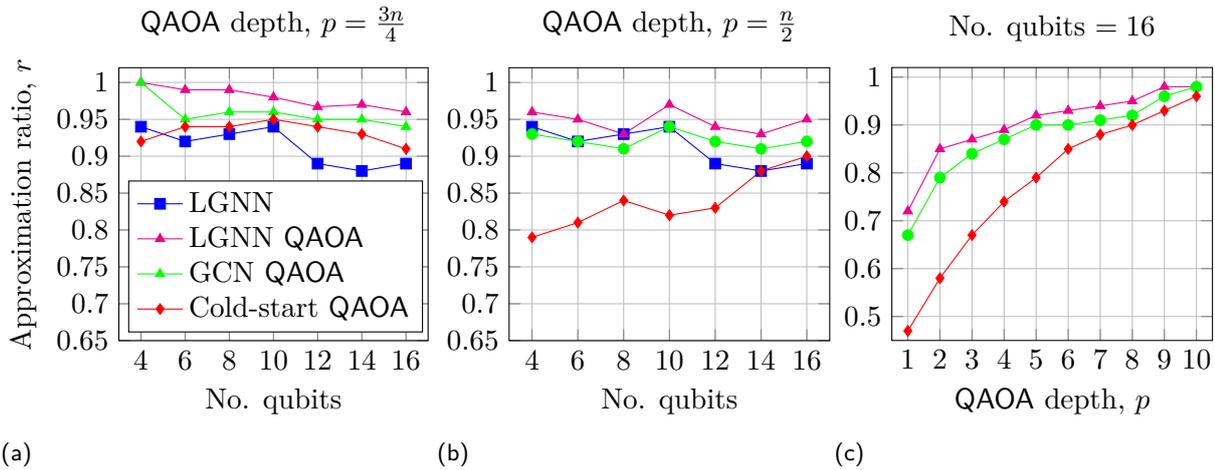
\begin{figure*}[!ht]
    \begin{subfigure}[t]{0.33\textwidth}
    \begin{tikzpicture}
    \begin{axis}[
        title = {$\QAOA$ depth, $p = \frac{3n}{4}$},
        xlabel={No. qubits},
        ylabel={Approximation ratio, $r$},
        ylabel near ticks,
        legend cell align={left},
        xmin=3, xmax=17,
        ymin=0.65, ymax= 1.02,
        xtick={ 4, 6, 8, 10, 12, 14, 16},
        ytick={0.6, 0.65, 0.70, 0.75, 0.80, 0.85, 0.90, 0.95, 1.00},
        legend pos=south west,
        ymajorgrids=true,
        grid   = major,
        height = 5.2cm,
        width  = 5.65cm
    ]
    \addplot[
        color=blue,
        mark=square*,
        ]
        coordinates {
     (4,   0.94)
     (6,   0.92)
     (8,   0.93)
     (10,   0.94)
     (12,   0.89)
     (14,   0.88)
     (16,   0.89)
    };
 
    \addplot[
        color=magenta,
        mark=triangle*,
        ]
        coordinates {
     (4,   1)
     (6,  0.99)
     (8,   0.99)
     (10,   0.98)
     (12,   0.967)
     (14,   0.97)
     (16,   0.96)
    };
    \addplot[
        color=green,
        mark=triangle*,
        ]
        coordinates {
     (4,   1)
     (6,  0.95)
     (8,   0.96)
     (10,   0.96)
     (12,   0.95)
     (14,   0.95)
     (16,   0.94)
    };
    \addplot[
        color=red,
        mark=diamond*,
        ]
        coordinates {
     (4,   0.92)
     (6,  0.94)
     (8,   0.94)
     (10,   0.95)
     (12,   0.94)
     (14,   0.93)
     (16,   0.91)
    };
    \legend{LGNN, LGNN $\QAOA$, GCN $\QAOA$ , Cold-start $\QAOA$}
    \end{axis}
    \end{tikzpicture}
    \caption{}
    \label{fig:gnn_maxcut_scores_a}
    \end{subfigure}
      \begin{subfigure}[t]{0.3\textwidth}
    \begin{tikzpicture}
    \begin{axis}[
        title = {$\QAOA$ depth, $p = \frac{n}{2}$},
        xlabel={No. qubits},
        ylabel={},
        ylabel near ticks,   
        legend cell align={left},
        xmin=3, xmax=17,
        ymin=0.65, ymax= 1.02,
        xtick={ 4, 6, 8, 10, 12, 14, 16},
        ytick={0.6, 0.65, 0.70, 0.75, 0.80, 0.85, 0.90, 0.95, 1.00},
        legend pos=south west,
        ymajorgrids=true,
        grid   = major,
        height = 5.2cm,
        width  = 5.8cm
    ]
    \addplot[
        color=blue,
        mark=square*,
        ]
        coordinates {
     (4,   0.94)
     (6,   0.92)
     (8,   0.93)
     (10,   0.94)
     (12,   0.89)
     (14,   0.88)
     (16,   0.89)
    };
    \addplot[
        color=magenta,
        mark=triangle*,
        ]
        coordinates {
     (4,   0.96)
     (6,  0.95)
     (8,   0.93)
     (10,   0.97)
     (12,   0.94)
     (14,   0.93)
     (16,   0.95)
    };
    \addplot[
        color=green,
        mark=*,
        ]
        coordinates {
     (4,   0.93)
     (6,  0.92)
     (8,   0.91)
     (10,   0.94)
     (12,   0.92)
     (14,   0.91)
     (16,   0.92)
    };
    \addplot[
        color=red,
        mark=diamond*,
        ]
        coordinates {
     (4,   0.79)
     (6,  0.81)
     (8,   0.84)
     (10,   0.82)
     (12,   0.83)
     (14,   0.88)
     (16,   0.90)
    };
    \end{axis}
    \end{tikzpicture}
    \caption{}
    \label{fig:gnn_maxcut_scores_b}
    \end{subfigure}
      \begin{subfigure}[t]{0.3\textwidth}
    \begin{tikzpicture}
    \begin{axis}[
        title = {No. qubits $=16$},
        xlabel={$\QAOA$ depth, $p$},
        ylabel={},
        ylabel near ticks,   
        legend cell align={left},
        xmin= 0.5, xmax=10.5,
        ymin=0.45, ymax= 1.02,
        xtick={ 1, 2, 3, 4,5,6,7,8,9, 10},
        ytick={0.4, 0.5, 0.6,0.70,  0.80,  0.90,  1.00},
        legend pos=south west,
        ymajorgrids=true,
        grid   = major,
        height = 5.2cm,
        width  = 5.8cm
    ]
    \addplot[
        color=magenta,
        mark=triangle*,
        ]
        coordinates {
     (1,   0.72)
     (2,  0.85)
     (3,   0.87)
     (4,   0.89)
     (5,   0.92)
     (6,   0.93)
     (7,   0.94)
     (8,   0.95)
     (9,   0.98)
     (10,   0.98)
    };
    \addplot[
        color=green,
        mark=*,
        ]
        coordinates {
     (1,   0.67)
     (2,  0.79)
     (3,   0.84)
     (4,   0.87)
     (5,   0.90)
     (6,   0.90)
     (7,   0.91)
     (8,   0.92)
     (9,   0.96)
     (10,   0.98)
    };
    \addplot[
        color=red,
        mark=diamond*,
        ]
        coordinates {
     (1,   0.47)
     (2,  0.58)
     (3,   0.67)
     (4,   0.74)
     (5,   0.79)
     (6,   0.85)
     (7,   0.88)
     (8,   0.90)
     (9,   0.93)
     (10,   0.96)
    };
    \end{axis}
    \end{tikzpicture}
    \caption{}
    \label{fig:gnn_maxcut_scores_c}
    \end{subfigure}
\caption{\textbf{Performance of the graph neural network on $3$-regular graphs for $\maxcut$.} We compare the $\maxcut$ approximation ratios achieved with LGNN/GCN initialisation versus cold-start (vanilla) $\QAOA$. We also plot the raw values outputted by the LGNN with a simple rounding scheme. In (a), the $\QAOA$ depth is set to be $p=3n/4$, while in (b), $p=n/2$. Each datapoint is generated via $1000$ runs of the LGNN/GCN on random instances of $3$-regular graphs, of the appropriate size to the number of qubits. Finally, in (c), we fix the number of qubits to be $16$ and plot each method as a function of the $\QAOA$ depth, demonstrating monotonic improvement as a function of $p$. Each datapoint is generated via $1000$ runs of the LGNN/GCN on random instances of $3$-regular graphs, of the appropriate size to the number of qubits.}
\label{fig:gnn_maxcut_numerics}
\end{figure*}

Next, \figref{fig:gnn_maxcut_numerics} compares the approximation ratio directly output by the GNN, against the GNN initialised solution which is then further optimised by $\QAOA$. To generate discrete $\maxcut$ solutions from the GNN probability vectors, we choose the same simple rounding scheme as~\cite{schuetz_combinatorial_2022}, assigning $\xbs_i^* = \text{int}(\boldsymbol{p}_i) \in \{0, 1\}$ for each node, $i$. We leave the incorporation and comparison of more complex rounding techniques to future work.

We first plot as function of qubit number (graph size), for $\QAOA$ depths scaling proportionally with the number of qubits ($p=3n/4$ in (a) and $p=n/2$ in (b)). We see that the GNN $\QAOA$ outperforms both the GNN and vanilla $\QAOA$ individually. This advantage is more pronounced at lower relative depth, but is diminished at higher depth. This is confirmed by \figref{fig:gnn_maxcut_numerics}(c) where we fix the qubit number at $16$ and plot the results as a function of $\QAOA$ depth. However, since the depth of quantum circuits is limited by decoherence in NISQ devices, the advantage of the GNN in the low-depth regime is promising. Finally, since we observe in these results that the LGNN appears to outperform the GCN architecture (due to the more complex message passing functionality), we opt to use the former primarily in the remainder of this work. However, for larger problem sizes, the LGNN may be less scalable~\cite{schuetz_combinatorial_2022}.

\subsection{Graph neural network versus SDP relaxations} \label{ssec:gnn_versus_gw_numerical}

Next, we benchmark the GNN against the SDP relaxation approach directly in~\figref{fig:gnn_versus_sdp}. We begin in~\figref{fig:gnn_vs_gw_solution_quality} by comparing the quality of the $\maxcut$ solution produced by the GW algorithm ($r_{\text{GW}}$) against the solution from the (line-)GNN ($r_{\text{GNN}}$). We observe comparable quality between the two methods (as remarked previously~\cite{schuetz_combinatorial_2022}) with the GNN generating solutions which are between $85$-$90\%$ the quality of the GW algorithm. However, we note that for these small graph instances, the training set becomes saturated as all possible $3$-regular graphs eventually appear. Naively increasing the size (say from $N_{\text{train}}=1000$ to $N_{\text{train}}=5000$) of training set with graphs of the \emph{same} size as the test case does not improve performance of the GNN dramatically. Therefore, in order to non-trivially increase the amount of training data, we include graphs of \emph{different} sizes (taking $N_{\text{train}} = 5000$ graphs which are $1\times$-$5\times$ the size of the test graph) in the training set. Doing so increases the performance ratio of GNN over GW to $\sim 95\%$, at the expense of a greater training time. However, if we then examine~\figref{fig:gnn_versus_gw_time_quantum} and~\figref{fig:gnn_versus_gw_time_classical}, then the tradeoff we are making becomes apparent. With a small sacrifice in solution quality, the GNN (once trained) is able to generate candidate solutions significantly faster than SDP relaxations and the GW algorithm.

Firstly, in~\figref{fig:gnn_versus_gw_time_quantum} and show that the inference time (time to produce a warm-started solution) is significantly higher with the SDP relaxation method than using a trained GNN. For this plot we focus graphs relevant to the $\QAOA$ problem sizes we study in this paper. Next, we show in~\figref{fig:gnn_versus_gw_time_classical} how the GNNs speed advantage enables the scalable production of warm-started solutions for $\QAOA$ into millions of variables, far beyond the capability of near term quantum devices. Specifically, we reproduce the results of~\cite{schuetz_combinatorial_2022} comparing the full time taken by the SDP relaxation and the GW algorithm against the GCN architecture (even \emph{including} training time) to solve $\maxcut$ (see~\cite{schuetz_combinatorial_2022} for details). The runtime of the GW algorithm is limited by the interior-point method used to solve SDPs which scales as $\widetilde{\mathcal{O}}(n^{3.5})$~\cite{schuetz_combinatorial_2022}. In~\secref{ssec:gnn_generalisation}, we take this even further and demonstrate how the GNN approach is able to generalise not only across test graph instances of the same size as in the training set, but also to \emph{larger} test graph instances, which is a feature clearly not possible via the relaxation method.

\begin{figure*}[!ht]
\begin{subfigure}[t]{0.45\textwidth}
\centering
\begin{tikzpicture}
\begin{axis}[
    xlabel={Epoch},
    ylabel={Approximation ratio, $r$},
    ylabel near ticks,
    xmin    =-5, xmax    =205,
    ymin    =0, ymax    = 1,
    xtick={0, 25, 50, 75, 100, 125, 150, 175, 200},
    ytick={0, 0.1, 0.2, 0.3, 0.4, 0.5, 0.6, 0.7, 0.8, 0.9, 1.0},
    legend pos=south east,
    legend cell align={left},
    ymajorgrids=true,
    grid =major,
    height = 6cm,
    width  = 8cm,
]
\addplot[
    color=blue,
    mark=*,
    ]
    coordinates {
(0,  0.20)
(25,	 0.41)
(50,	 0.61)
(75,	  0.85)
(100, 0.92)
(125,	 0.94)
(150,	 0.95)
(175,	 0.95)
(200,	 0.95)

};

\addplot[
    color=orange,
    mark=triangle*,
    ]
    coordinates {
(0,  0.40)
(25,	 0.65)
(50,	 0.75)
(75,	  0.90)
(100, 0.92)
(125,	 0.94)
(150,	 0.95)
(175,	 0.95)
(200,	 0.95)};

\addplot[
    color=green,
    mark=square*,
    ]
    coordinates {
(0,  0.37)
(25,	 0.58)
(50,	 0.72)
(75,	  0.87)
(100, 0.93)
(125,	 0.94)
(150,	 0.95)
(175,	 0.95)
(200,	 0.95)
};
\addplot[
    color=red,
    mark=diamond*,
    ]
    coordinates {
(0,  0.02)
(25,	 0.23)
(50,	 0.35)
(75,	  0.60)
(100, 0.72)
(125,	 0.77)
(150,	 0.87)
(175,	 0.92)
(200,	 0.92)
};

\legend{TQA $\QAOA$, Relax $\QAOA$, LGNN $\QAOA$, Cold-start $\QAOA$}

\end{axis}
\end{tikzpicture}
\caption{}
\end{subfigure}
\begin{subfigure}[t]{0.45\textwidth}
\centering
\begin{tikzpicture}
\begin{axis}[
    xlabel={$\QAOA$ depth $p$},
    ylabel={Approximation ratio $r$},
        ylabel near ticks,
    xmin=0.5, xmax=5.5,
    ymin=0.65, ymax= 1.02,
    xtick={0, 1, 2, 3, 4, 5, 6},
    ytick={0.6, 0.65, 0.7, 0.75,  0.8,0.85, 0.9, 0.95, 1.0},
    legend pos=south east,
    legend cell align={left},
    ymajorgrids=true,
    grid   = major,
    height = 6cm,
    width  = 8cm,
]
\addplot[
    color=blue,
    mark=*,
    ]
    coordinates {
(1,  0.80)
(2,	 0.87)
(3,	  0.90)
(4, 0.92)
(5,	 0.98)
};

\addplot[
    color=orange,
    mark=triangle*,
    ]
    coordinates {
(1,  0.89 )
(2,	0.91)
(3,	 0.97)
(4,	1.0)
(5,	1.0)
};

\addplot[
    color=green,
    mark=square*,
    ]
    coordinates {
(1, 0.87)
(2, 0.88)
(3,  0.95)
(4, 0.95)
(5,  0.98)
};
\addplot[
    color=red,
    mark=diamond*,
    ]
    coordinates {
(1, 0.68)
(2, 0.82)
(3,  0.84)
(4, 0.89)
(5,  0.96)
};

\end{axis}
\end{tikzpicture}
\caption{}
\end{subfigure}
\caption{\textbf{Comparison of all initialisation techniques}. We use $3$-regular graphs over $8$ qubits and for TQA we choose $\delta t= 0.75$ as in~\cite{sack_quantum_2021}. (a) Convergence of initialisations as a function of training iteration. The depth of QAOA is fixed to $5$. (b) Comparison of initialisations as a function of $\QAOA$ depth $p$. We plot the average for each method over $10$ runs.}
\label{fig:comparing_all_inits}
\end{figure*}
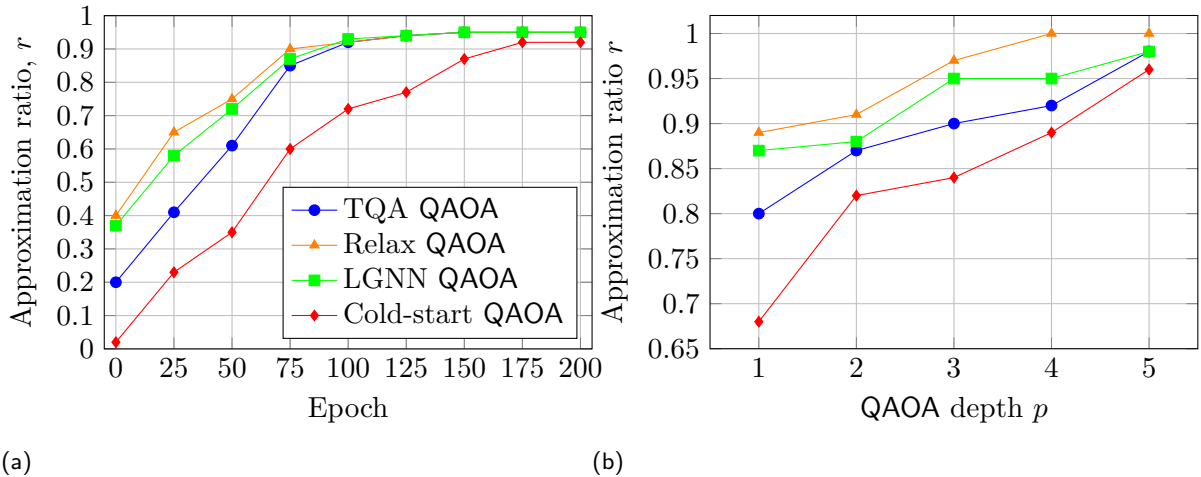

As a final comparison, we compare the GNN initialisation technique against all other techniques in~\figref{fig:comparing_all_inits}. We compare against the warm-starting technique using relaxations of~\cite{egger_warm-starting_2021} (`Warm-start'), and the Trotterised quantum annealing (`TQA') based approach of~\cite{sack_quantum_2021}, as a function of depth and training epochs.

\begin{figure*}[!ht]
\begin{subfigure}[t]{0.3\textwidth}
 \begin{tikzpicture}
\begin{axis}[
    xlabel={Graph Size (No. Nodes)},
    ylabel={$r_{\text{GNN}}$/$r_{\text{GW}}$},
    ylabel near ticks,
    xlabel near ticks,
    xmin=3.5, xmax=16.5,
    ymin=0.85, ymax= 0.98,
    xtick={0, 2, 4, 6, 8, 10, 12, 14, 16},
    ytick={0.86, 0.88, 0.9, 0.92, 0.94, 0.96, 0.98},
    ymajorgrids=true,
    legend pos=south east,
    grid =major,
    height = 5cm,
    width  = 5.25cm
]
\addplot[
    color=blue,
    mark=*,
    ]
    coordinates {
(4, 0.89)
(6,  0.92)
(8,	  0.92)
(10,	 0.94)
(12,	 0.91)
(14,	 0.92)
(16,	 0.91)

};

\addplot[
    color=green,
    mark=*,
    ]
    coordinates {
(4, 0.94)
(6,  0.96)
(8,	  0.95)
(10,	 0.96)
(12,	 0.94)
(14,	 0.95)
(16,	 0.96)

};
\legend{$N_{\text{train}}=1000$, $N_{\text{train}}=5000$}
\end{axis}
\end{tikzpicture}
\caption{}
\label{fig:gnn_vs_gw_solution_quality}
\end{subfigure}
\begin{subfigure}[t]{0.3\textwidth}
\begin{tikzpicture}
\begin{axis}[
    xlabel={No. Qubits},
    ylabel={Time (ms)},
    ylabel near ticks,
    xlabel near ticks,
    xmin=3, xmax=21,
    ymin=0, ymax= 300,
    xtick={ 4, 8, 12, 16, 20},
    ytick={0, 40, 80, 120, 160, 200, 240, 280},
    legend pos=north west,
    legend cell align={left},
    ymajorgrids=true,
    grid =major,
        height = 5cm,
        width  = 5.25cm
]

\addplot[
    color=orange,
    mark=*,
    ]
    coordinates {
(4,  20)
(6,	 29)
(8,	  56)
(10, 77)
(12,  99)
(14,  121)
(16,	 175)
(18,	 219)
(20, 260)

};

\addplot[
    color=blue,
    mark=*,
    ]
    coordinates {
(4,  52)
(6,	 55)
(8,	  56)
(10, 59)
(12, 61)
(14,  67)
(16, 69)
(18, 70)
(20, 72)

};
\legend{Relaxation, LGNN}
\end{axis}
\end{tikzpicture}
\caption{}
\label{fig:gnn_versus_gw_time_quantum}
\end{subfigure}
\begin{subfigure}[t]{0.3\textwidth}
\begin{tikzpicture}
\begin{axis}[
    xmode=log,
    ymode=log,
    xlabel={No. Nodes},
    ylabel={Time (s)},
    ylabel near ticks,
    xlabel near ticks,
    xmin=0, xmax=2100100,
    ymin=0, ymax= 5000,
    xtick={ 100, 1000, 10000, 100000, 1000000},
    ytick={0, 10, 100, 1000},
    legend pos=south east,
    legend cell align={left},
    ymajorgrids=true,
    grid =major,
        height = 5cm,
        width  = 5.5cm
]
\addplot[
    color=cyan,
    mark=*,
    ]
    coordinates {
(100,  46)
(125,	 245)
(250,	  825)
(375, 2800)

};

\addplot[
    color=magenta,
    mark=*,
    ]
    coordinates {
(100,  6.8)
(125,	 7.4)
(250,	  8.1)
(375, 8.9)
(750, 13.7)
(1000,  51.3)
(4000, 53.6)
(7000, 57.7)
(10000, 58.6)
(30000, 63.2)
(75000, 66.8)
(100000, 90.7)
(400000, 112.5)
(700000, 487.4)
(1000000, 767.7)
};
\legend{GW, GCN}
\end{axis}
\end{tikzpicture}
\caption{}
\label{fig:gnn_versus_gw_time_classical}
\end{subfigure}
\caption{\textbf{Time versus quality tradeoff between GNN versus relaxation initialisation methods.} (a) $\maxcut$ approximation ratios generated by GNN and the continuous relaxation (without $\QAOA$), as a function of graph size. $r_j$ is the approximation ratio generated by method $j\in \{\text{GW}, \text{GNN}\}$. We use a simple rounding technique to generate the discrete values from the soft outcomes from the (line-)GNN (see main text) and plot results for $1000$ and $5000$ training graphs. (b) Comparison of time to produce a warm-start taken by relaxation initialisation versus (line-)GNN initialisation as a function of qubit number, averaged over $10$ random graph instances. The GNN enables much faster inference for $\maxcut$. This does not include pre-training time for the GNN, but as an example, training on $1000$ graphs for $18$ qubits takes only $6$ minutes for the LGNN. Finally, (c) reproduces results taken from~\cite{schuetz_combinatorial_2022} demonstrating the scalability of the GNN initialisation (for the GCN) over solving the SDP relaxation with the GW algorithm. The GNN can provide warm-starts for $\QAOA$ on graphs with millions of nodes. Note, these plots do not include $\QAOA$ runtime after the warm-start.}
\label{fig:gnn_versus_sdp}
\end{figure*}
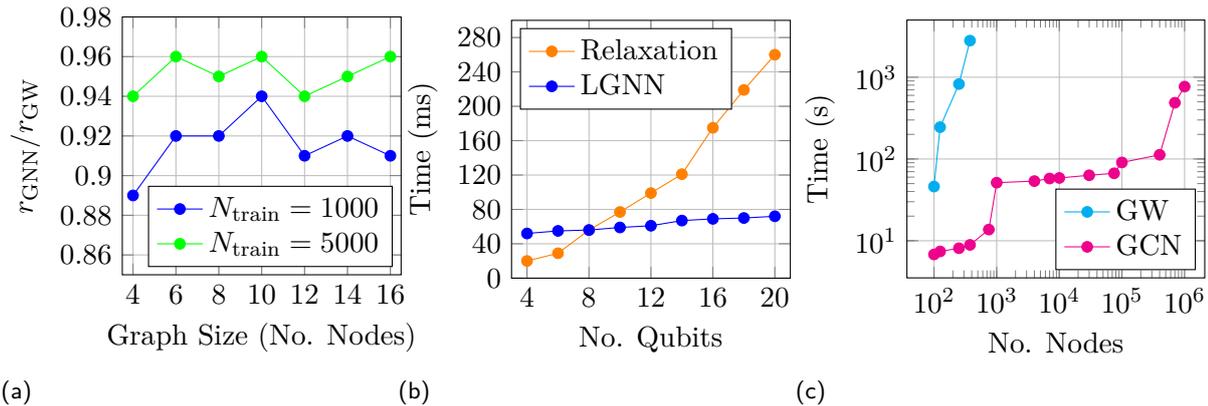

\subsection{Generalisation capabilities of GNNs} \label{ssec:gnn_generalisation}

A key feature of using neural networks for certain tasks is capability to \emph{generalise}. This generalisation ability is one of the driving motivations behind machine learning, and tests the ability of an algorithm to actually \emph{learn} (rather than just memorise a solution). Similarly, we can test the generalisation ability of GNNs in warm-starting the $\QAOA$. To do so, we train instantiations of GNNs on small graph instances, and then directly apply them to larger test instances. We test this for examples of between $6$ to $14$ nodes in~\tabref{table:gnn_generalisation}. Here, we see this generalisation feature directly, as the GNN is capable of performing well on graphs larger than those in the training set. Note a related generalisation behaviour was demonstrated via meta-learning~\cite{verdon_learning_2019, sauvage_flip_2021} for the \emph{parameters} of a variational circuit. The work of~\cite{verdon_learning_2019} utilises recurrent neural networks (we revisit this strategy in~\secref{ssec:ml_for_qaoa}) for training the $\QAOA$, but generalisation in this case was possible due to the structure of the algorithm and parameter concentration~\cite{akshay_parameter_2021}. The FLexible Initializer
for arbitrarily-sized Parametrized quantum circuits (FLIP)~\cite{sauvage_flip_2021}, also has related parameters generalisation capabilities, which would be interesting to compare and incorporate with warm-starting initialisers in future work.

The rows in~\tabref{table:gnn_generalisation} correspond to the graph size on which the model was trained, and the columns correspond to the graph size on which the GNN was then tested. For example, both training and testing on graphs with $10$ nodes gives an approximation ratio of $\approx 0.93$, whereas if we train the GNN using only graphs of $6$ nodes, there is no drop in performance. In contrast, if we reduce from $14$ to $6$ training nodes, we only see a drop in the approximation ratio of $7\%$. Again, we mention the limitation to small problem sizes due to the $\QAOA$ circuit simulation overhead. As with the discussions above, it has been observed that GNNs trained on $30$ node graphs have the ability to generalise to $300$ nodes and larger~\cite{yao_experimental_2019, schuetz_combinatorial_2022}. Note that this generalisation is the reverse situation to that in \figref{fig:gnn_vs_gw_solution_quality}, where we include larger graphs in the \emph{training} set to improve solution quality in smaller graphs, rather than including larger graphs in the \emph{test} set, as we do here.

\begin{table}
\begin{adjustbox}{width=\columnwidth,center}
\begin{tabular}{||c|c|c|c|c||}
\hline
\backslashbox{Train size}{Test size} & 8 & 10 & 12  & 14 \\ 
\hline
6 & 0.91 & 0.93 & 0.89 & 0.89 \\
\hline
8 & \textbf{0.93} & 0.92 & 0.89 & 0.91 \\
\hline
10 &  \cellcolor{blue!25}&  \textbf{0.93} & 0.90& 0.89\\
\hline
12 &  \cellcolor{blue!25} &   \cellcolor{blue!25} & \textbf{0.90} & 0.89 \\
\hline
14 &  \cellcolor{blue!25} &   \cellcolor{blue!25} & \cellcolor{blue!25} & \textbf{0.96}\\
\hline
\end{tabular}
\end{adjustbox}
\caption{Value of approximation ratio, $r$, as a function of training and test graph size. In each case, both the train and test set consists of $1000$ random graphs of the appropriate sizes. \textbf{Bold} indicates the instances where the train graph size is the same as the test size.}
\label{table:gnn_generalisation}
\end{table}

\section{Optimisation of the \texorpdfstring{$\QAOA$}{}} \label{sec:qaoa_optimisation}

Now, we move to the second focus of this work, which is a comparison between a wide range of different optimisers which can be used to train the $\QAOA$ when solving the $\maxcut$ problem. A large variety of optimisers for VQAs have been proposed in the literature, and each has their own respective advantages and disadvantages when solving a particular problem. Due to the hybrid quantum-classical nature of these algorithms, many of optimisation techniques have been taken directly from classical literature in, for example, the training of classical neural networks. However, due to the need to improve the performance of near term quantum algorithms, there has also been much effort put into the discovery of \emph{quantum-aware} optimisers, which may include many non-classical hyperparameters including, for example, the number of measurement shots to be taken to compute quantities of interest from quantum states~\cite{sweke_stochastic_2020, kubler_adaptive_2020}. In the following sections, we implement and compare a number of these optimisers. We begin by evaluating gradient-based and gradient-free optimisers in~\secref{sec:qaoa_optimisation}. We then compare quantum and classical methods for simultaneous perturbation stochastic optimisation in~\secref{ssec:spsa_optimisers}, which typically have lower overheads than the gradient-free or gradient-based optimisers since all parameters are updated in a single optimisation step, as opposed to parameter-wise updates. Finally, we then implement some neural network based optimisers in~\secref{sec:nn_qaoa_training} which operate via reinforcement learning and the meta-learning optimiser mentioned above. In all of the above cases, we use vanilla stochastic gradient descent optimisation as a benchmark.

\subsection{Gradient-based versus gradient-free optimisation} \label{sec:gradient_free_vs_gradient} 

For a given parameterised quantum circuit instance, equipped with parameters at iteration (epoch), $t$, $\paramtheta^{t}$, the parameters at iteration $t+1$ are given by an update rule:
\begin{equation}\label{eqn:opt_update_rule}
    \paramtheta^{t+1} = \paramtheta^{t} - \Delta \mathsf{C}(\paramtheta^{t})
\end{equation}
$\Delta \mathsf{C}(\paramtheta^{t})$ is the update rule, which contains information about the cost function to be optimised at epoch $t$, $\mathsf{C}(\paramtheta^{t})$. In gradient-based optimisers, the update contains information about the gradients of $\mathsf{C}(\paramtheta^{t})$ with respect to the parameters, $\paramtheta^t$. In contrast, gradient-free (or zeroth order) optimisation methods use only information about $\mathsf{C}(\paramtheta^{t})$ itself. One may also incorporate second-order derivative information also, which tend to outperform the previous methods, but are typically more expensive as a result. In the following, we test techniques which fall into all of these categories, for a range of qubit numbers and $\QAOA$ depths. 

A general form of this update rule when incorporating gradient information can be written as:
\begin{equation}\label{eqn:opt_update_rule_natural_gradient}
    \paramtheta^{t+1} = \paramtheta^{t} -  \eta(t, \paramtheta) \boldsymbol{g}(\paramtheta)^{-1}\nabla \mathsf{C}(\paramtheta^{t})
\end{equation}
where $\eta(t, \paramtheta)$ is a \emph{learning rate}, which determines the speed of convergence, and may depend on the previous parameters, $\paramtheta$ and $t$. The quantity $\boldsymbol{g}(\paramtheta) \in \mathbb{R}^{d \times d}$ is a \emph{metric tensor}, which incorporates information about the parameter landscape. This tensor can be the classical, or quantum Fisher information (QFI) for example. In the case of the latter, the elements of $\boldsymbol{g}(\paramtheta)$ when dealing with a parameterised state, $\ket{\psi_{\paramtheta}}$ are given by:
\begin{multline}
    \boldsymbol{g}_{ij}(\paramtheta) := \text{Re}\bigg\{\bigg\<\frac{\partial \psi_{\paramtheta}}{\partial \paramtheta_i}\bigg|\frac{\partial \psi_{\paramtheta}}{\partial \paramtheta_j}\bigg\> \\
    \left.- \bigg\<\frac{\partial \psi_{\paramtheta}}{\partial \paramtheta_i}\bigg|{\psi_{\paramtheta}}\bigg\>\bigg\<\psi_{\paramtheta}\bigg|\frac{\partial \psi_{\paramtheta}}{\partial \paramtheta_j}\bigg\>\right\}
\end{multline}
In this form, the gradient update~\eqnref{eqn:opt_update_rule_natural_gradient} updates the parameters according to the quantum \emph{natural} gradient (QNG)~\cite{stokes_quantum_2020}.

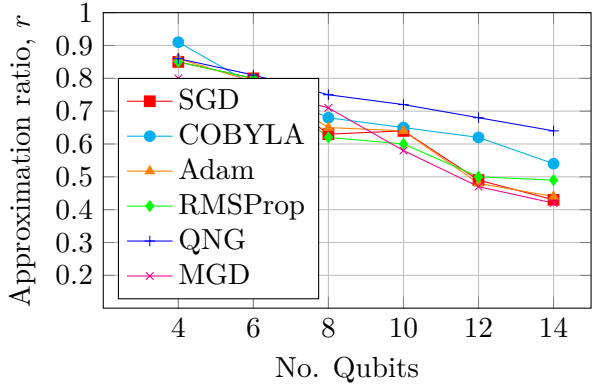
\begin{figure}[!ht]
\centering
\begin{tikzpicture}
\begin{axis}[
    xlabel={No. Qubits},
    ylabel={Approximation ratio, $r$},
    ylabel near ticks,
    xmin=2, xmax=15,
    ymin=0.1, ymax= 1,
    xtick={0, 4, 6, 8, 10, 12, 14},
    ytick={0.2, 0.3, 0.4, 0.5, 0.6,  0.7, 0.8, 0.9 , 1.0},
    legend pos=south west,
    legend cell align={left},
    ymajorgrids=true,
    grid =major,
    height = 5.5cm,
    width  = 8cm,
]
\addplot[
    color=red,
    mark=square*,
    ]
    coordinates {
(4,  0.85)
(6,	 0.80)
(8,	  0.63)
(10, 0.64)
(12,  0.49)
(14,  0.43)

};
\addplot[
    color=cyan,
    mark=*,
    ]
    coordinates {
(4,  0.91)
(6,	 0.78)
(8,	  0.68)
(10, 0.65)
(12,  0.62)
(14,  0.54)

};

\addplot[
    color=orange,
    mark=triangle*,
    ]
    coordinates {
(4,  0.86)
(6,	 0.79)
(8,	  0.65)
(10, 0.64)
(12,  0.48)
(14,  0.44)

};

\addplot[
    color=green,
    mark=diamond*,
    ]
    coordinates {
(4,  0.85)
(6,	 0.80)
(8,	  0.62)
(10, 0.60)
(12,  0.50)
(14,  0.49)
};
\addplot[
    color=blue,
    mark=+,
    ]
    coordinates {
(4,  0.86)
(6,	 0.81)
(8,	  0.75)
(10, 0.72)
(12,  0.68)
(14,  0.64)
};

\addplot[
    color=magenta,
    mark=x,
    ]
    coordinates {
(4,  0.80)
(6,	 0.77)
(8,	  0.71)
(10, 0.58)
(12,  0.47)
(14,  0.42)
};

\legend{SGD, COBYLA, Adam, RMSProp, QNG, MGD}
\end{axis}
\end{tikzpicture}
\caption{\textbf{Gradient-based versus gradient-free optimisers}. We set the depth of $\QAOA$ depth to $p=4$ and vary the number of qubits. All optimisers have been run $10$ times and average values have been plotted.}
\label{fig:optimiser_comparison}
\end{figure}

If we further simplify by taking $\boldsymbol{g} = \mathds{1}$ to be the identity and choosing different functions for $\eta(\paramtheta, t)$, we recover many popular optimisation routines such as Adam~\cite{kingma_adam_2015} or Adadelta~\cite{zeiler_adadelta_2012}, which incorporates notions such as \emph{momentum} into the update rule, and makes the learning rate time-dependent. Such behaviour is desired to, for example, allow the parameters to make large steps at the beginning of optimisation (when far from the target), and take smaller steps towards the latter stage when one is close to the optimal solution. The simplest form of gradient descent is \emph{vanilla}, which takes $\eta(\paramtheta, t) := \eta$ to be a constant. The `stochastic' versions of gradient descent use an approximation of the cost gradient computed with only a few training examples.
In~\figref{fig:optimiser_comparison}, we begin by comparing some examples of the above gradient-based optimisation rule (specifically using QNG, Adam and RMSProp) to a gradient-free method (COBYLA)~\cite{powell_direct_1994}. We also add a method known as \emph{model} gradient descent (MGD) which is a gradient-based method introduced by~\cite{harrigan_quantum_2021} that involves quadratic model fitting as a gradient estimation (see Appendix A of~\cite{sung_using_2020} for pseudocode). The results are shown in~\figref{fig:optimiser_comparison} for $\maxcut$ on $3$ regular graphs for up to $14$ qubits. We observe that optimisation using the QNG outperforms other methods, however it does so with a large resource requirement, which is needed to compute the quantum fisher information (QFI) using quantum circuit evaluations.

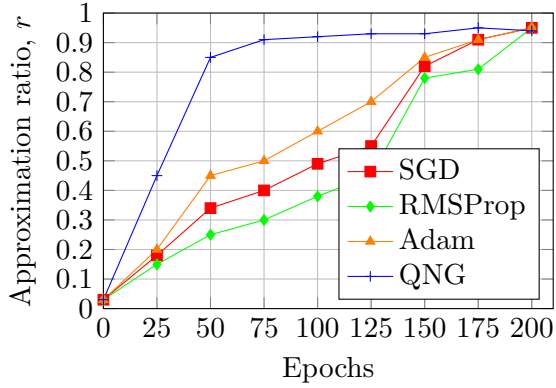
\begin{figure}[!ht]
\centering
\begin{tikzpicture}
\begin{axis}[
    xlabel={Epochs},
    ylabel={Approximation ratio, $r$},
    ylabel near ticks,
    xmin    =0, xmax    =210,
    ymin    =0, ymax    = 1,
    xtick={0, 25, 50, 75, 100, 125, 150, 175, 200},
    ytick={0, 0.1, 0.2, 0.3, 0.4, 0.5, 0.6, 0.7, 0.8, 0.9, 1.0},
    legend pos  =south east,
    legend cell align={left},
    ymajorgrids =true,
    grid        = major,
    height      = 5.5cm,
    width       = 7.5cm,
]
\addplot[
    color=red,
    mark=square*,
    ]
    coordinates {
(0,  0.03 )    
(25,  0.18 )
(50,	0.34)
(75,	 0.40)
(100,	0.49)
(125,	0.55)
(150,  0.82)
(175,	0.91)
(200,	 0.95)

};

\addplot[
    color=green,
    mark=diamond*,
    ]
    coordinates {
(0,  0.03 )    
(25,  0.15 )
(50,	0.25)
(75,	 0.30)
(100,	0.38)
(125,	0.45)
(150,  0.78 )
(175,	0.81)
(200,	 0.95)

};
\addplot[
    color=orange,
    mark=triangle*,
    ]
    coordinates {
(0,  0.03 )    
(25,  0.20 )
(50,	0.45)
(75,	 0.50)
(100,	0.60)
(125,	0.70)
(150,  0.85 )
(175,	0.91)
(200,	 0.95)

};
\addplot[
    color=blue,
    mark=+,
    ]
    coordinates {
(0,  0.03 )    
(25,  0.45 )
(50,	0.85)
(75,	 0.91)
(100,	0.92)
(125,	0.93)
(150,  0.93 )
(175,	0.95)
(200,	 0.94)

};

\legend{SGD, RMSProp, Adam , QNG}
\end{axis}
\end{tikzpicture}
\caption{Comparison of optimisers relative to convergence speed. We fix $\QAOA$ depth at $p=6$ with $14$ qubits.}
\label{fig:qng_versus_adam_rmsprop}
\end{figure}

We next examine the convergence speed of the `quantum-aware' QNG optimiser, versus the standard Adam and RMSProp in~\figref{fig:qng_versus_adam_rmsprop}. Again, QNG significantly reduces convergence time, but again at the expense of being a more computationally taxing optimisation method~\cite{stokes_quantum_2020}.

\subsection{Simultaneous perturbation stochastic approximation optimisation} \label{ssec:spsa_optimisers}

From the above, using the QNG as an optimisation routine is very effective, but it has a large computational burden due to the evaluation of the quantum Fisher information. A strategy to bypass this inefficiency was proposed by~\cite{gacon_simultaneous_2021}, who suggested combining the QNG gradient with the \emph{simultaneous perturbation stochastic approximation} (SPSA) algorithm. This algorithm is an efficient method to bypass the linear scaling in the number of parameters using the standard parameter shift rule~\cite{mitarai_quantum_2018, schuld_evaluating_2019}, to compute quantum gradients. For example, in the expression~\eqnref{eqn:opt_update_rule_natural_gradient} when restricted to vanilla gradient descent, one gradient term must be computed for each of the $d$ ($d=2p$ in the case of the $\QAOA$) parameters. In contrast, SPSA  approximates the entire gradient vector by choosing a \emph{random} direction in parameter space and estimating the gradient in this direction using, for example, a finite difference method. This requires a constant amount of computation relative to the number of parameters. To incorporate the quantum Fisher information,~\cite{gacon_simultaneous_2021} to SPSA actually uplifts a \emph{second order} version of SPSA (called $2$-SPSA), which exploits the Hessian of the cost function to be optimised. The update rules for $1$-,  $2$-SPSA and QN-SPSA are given by:
\begin{align}
    \paramtheta^{t+1} &  = \paramtheta^t - \eta\times
    \begin{cases}
     \widetilde{\nabla} \mathsf{C}(\paramtheta^t) &\quad 1\text{-SPSA}\\
     \widetilde{H}^{-1}(\paramtheta^t)\widetilde{\nabla} \mathsf{C}(\paramtheta^t) &\quad 2\text{-SPSA}\\ \widetilde{\boldsymbol{g}}^{-1}(\paramtheta^t)\widetilde{\nabla}\mathsf{C}(\paramtheta^t) &\quad \text{QN-SPSA}
        \end{cases}
\end{align}
where stochastic approximation to the Hessian\footnote{The actual quantity used in \cite{gacon_simultaneous_2021} is a weighted quantity combining these approximations at all previous time steps.}, $\widetilde{H}(\paramtheta)$, and the quantum Fisher information are given by:
\begin{equation}
    \widetilde{H}^{t} := -\frac{1}{2}\frac{\delta \mathsf{C}}{2\epsilon^2}\frac{\Delta_1^t(\Delta_2^t)^T+\Delta_2^t(\Delta_1^t)^T}{2}
    \end{equation}
\begin{multline}
    \delta \mathsf{C} := \mathsf{C}(\paramtheta^t + \epsilon\Delta_1^t + \epsilon \Delta_2^t)
    - \mathsf{C}(\paramtheta^t + \epsilon\Delta_1^t )\\
    - \mathsf{C}(\paramtheta^t - \epsilon\Delta_1^t + \epsilon \Delta_2^t)
    + \mathsf{C}(\paramtheta^t - \epsilon\Delta_1^t)
\end{multline}
and
\begin{equation}
    \widetilde{\boldsymbol{g}}^{t} = -\frac{1}{2}\frac{\delta F}{2\epsilon^2}\frac{\Delta_1^t(\Delta_2^t)^T+\Delta_2^t(\Delta_1^t)^T}{2}
    \end{equation}
\begin{multline}
    \delta F := F(\paramtheta^t, \paramtheta^t + \epsilon\Delta_1^t + \epsilon \Delta_2^t)
    - F(\paramtheta^t, \paramtheta^t + \epsilon\Delta_1^t )\\
    - F(\paramtheta^t, \paramtheta^t - \epsilon\Delta_1^t + \epsilon \Delta_2^t)
    + F(\paramtheta^t, \paramtheta^t - \epsilon\Delta_1^t)
\end{multline}
respectively. Here, $F(\paramtheta, \paramtheta + \boldsymbol{\alpha}) = |\langle{\psi_{\paramtheta}}|\psi_{\paramtheta+ \boldsymbol{\alpha}}\rangle|^2$ is the fidelity between the parameterised state prepared with angles, $\paramtheta$, and a `shifted' version with angles $\paramtheta+ \boldsymbol{\alpha}$. The quantities $\Delta_1, \Delta_2$ are uniformly random vectors sampled over $\{-1, 1\}^d$. Also, $\epsilon$ is a small constant arising from finite differencing approximations of the gradient, for example, we approximate the true gradient, $\nabla \mathsf{C}(\paramtheta)$ by $\widetilde{\nabla} \mathsf{C}(\paramtheta)$, given by:
\begin{equation}
    \widetilde{\nabla} \mathsf{C}(\paramtheta^t) := \frac{\mathsf{C}(\paramtheta^t + \epsilon \Delta^t) - \mathsf{C}(\paramtheta^t - \epsilon \Delta^t)}{2 \epsilon} \Delta^t
\end{equation}

\begin{figure*}[!ht]
\begin{subfigure}[b]{0.3\textwidth}
\centering
\begin{tikzpicture}
\begin{axis}[
    xlabel={$\QAOA$ depth, $p$},
    ylabel={Approximation ratio, $r$},
    xlabel near ticks,
    ylabel near ticks,
    xmin=0.5, xmax=6.5,
    ymin=0.1, ymax= 1,
    xtick={0, 1, 2, 3, 4, 5, 6, 7, 8},
    ytick={0, 0.2, 0.4, 0.6, 0.8, 1.0},
    legend pos=south east,
    legend cell align={left},
    ymajorgrids=true,
    grid =major,
    height = 5cm,
    width  = 5.5cm,
]
\addplot[
    color=red,
    mark=square*,
    ]
    coordinates {
(1, 0.18)
(2, 0.42)
(3,  0.48)
(4, 0.63)
(5,  0.83)
(6,  0.93)
};

\addplot[
    color=green,
    mark=diamond*,
    ]
    coordinates {
(1, 0.20)
(2, 0.35)
(3,  0.38)
(4, 0.67)
(5,  0.86)
(6,  0.93)
};

\addplot[
    color=magenta,
    mark=*,
    ]
    coordinates {
(1, 0.44)
(2, 0.61)
(3,  0.67)
(4, 0.77)
(5,  0.89)
(6,  0.933)
};
\addplot[
    color=cyan,
    mark=triangle*,
    ]
    coordinates {
(1, 0.64)
(2, 0.72)
(3,  0.77)
(4, 0.80)
(5,  0.90)
(6,  0.94)
};

\end{axis}
\end{tikzpicture}
\caption{}
\end{subfigure}
\begin{subfigure}[b]{0.3\textwidth}
\centering
\begin{tikzpicture}
\begin{axis}[
    xlabel={No. Qubits},
    xlabel near ticks,
    ylabel near ticks,
    xmin=5.5, xmax=17,
    ymin=0.5, ymax= 1,
    xtick={6, 8, 10, 12, 14, 16},
    ytick={0, 0.2, 0.4, 0.6, 0.8, 1.0},
    legend pos=south east,
    legend cell align={left},
    ymajorgrids=true,
    grid =major,
    height = 5cm,
    width  = 5.5cm,
]
\addplot[
    color=red,
    mark=square*,
    ]
    coordinates {
(6, 0.96)
(8, 0.85)
(10,  0.76)
(12, 0.66)
(14,  0.63)
(16,  0.53)
};

\addplot[
    color=green,
    mark=diamond*,
    ]
    coordinates {
(6, 0.98)
(8, 0.84)
(10,  0.77)
(12, 0.68)
(14,  0.62)
(16,  0.56)
};

\addplot[
    color=magenta,
    mark=*,
    ]
    coordinates {
(6, 0.98)
(8, 0.90)
(10,  0.83)
(12, 0.77)
(14,  0.68)
(16,  0.62)
};
\addplot[
    color=cyan,
    mark=triangle*,
    ]
    coordinates {
(6, 0.99)
(8, 0.91)
(10,  0.86)
(12, 0.77)
(14,  0.72)
(16,  0.67)
};

\end{axis}
\end{tikzpicture}
\caption{}
\end{subfigure}
\begin{subfigure}[b]{0.3\textwidth}
\centering
\begin{tikzpicture}
\begin{axis}[
    xlabel={Epochs},
    xlabel near ticks,
    ylabel near ticks,
    xmin    =0, xmax    =210,
    ymin    =0, ymax    = 1,
    xtick={0, 25, 50, 75, 100, 125, 150, 175, 200},
    ytick={0, 0.2, 0.4, 0.6, 0.8, 1.0},
    legend pos  = south east,
    legend cell align={left},
    ymajorgrids =true,
    grid        = major,
  height = 5cm,
    width  = 6.5cm,
]
\addplot[
    color=red,
    mark=square*,
    ]
    coordinates {
(0,  0.03 )    
(25,  0.20 )
(50,	0.35)
(75,	 0.43)
(100,	0.51)
(125,	0.66)
(150,  0.83)
(175,	0.89)
(200,	 0.94)

};

\addplot[
    color=green,
    mark=diamond*,
    ]
    coordinates {
(0,  0.03 )    
(25,  0.25 )
(50,	0.44)
(75,	 0.51)
(100,	0.58)
(125,	0.62)
(150,  0.77 )
(175,	0.89)
(200,	 0.94)

};

\addplot[
    color=magenta,
    mark=*,
    ]
    coordinates {
(0,  0.03 )    
(25,  0.27 )
(50,	0.60)
(75,	 0.69)
(100,	0.76)
(125,	0.79)
(150,  0.88 )
(175,	0.93)
(200,	 0.94)

};

\addplot[
    color=cyan,
    mark=triangle*,
    ]
    coordinates {
(0,  0.03 )    
(25,  0.35 )
(50,	0.73)
(75,	 0.85)
(100,	0.88)
(125,	0.93)
(150,  0.94 )
(175,	0.94)
(200,	 0.94)

};

\legend{SGD, 1-SPSA, 2-SPSA, QN-SPSA}

\end{axis}
\end{tikzpicture}
\caption{}
\end{subfigure}
\caption{\textbf{Simultaneous perturbation stochastic approximation ($1$-, $2$- and QN-SPSA) for $\maxcut$ with $\QAOA$.} Plots show approximation ratio as a function of (a) depth (qubit number fixed at $10$), (b) number of qubits ($\QAOA$ depth is fixed to $p=5$) and (c) training iteration (qubit number and $\QAOA$ depth fixed to $10$ and $7$ respectively). In all cases, the average is taken over $10$ independent optimisation runs.}
\label{fig:spsa_method_comparison}
\end{figure*}
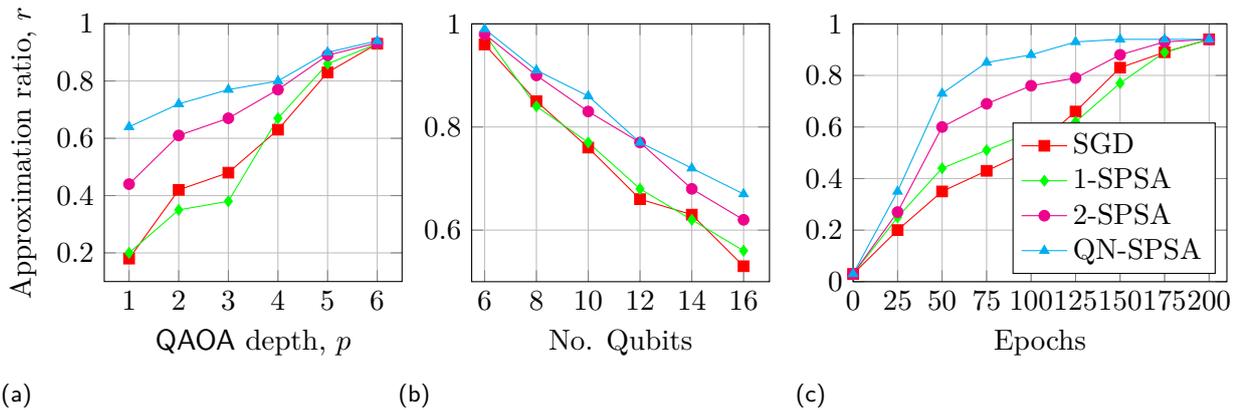

We compare all these three perturbation methods in~\figref{fig:spsa_method_comparison} against SGD once again, with a fixed learning rate of $\eta=0.01$. Notice that SGD performs comparably to $1$-SPSA, but with the expense of more cost function evaluations.

\subsection{Neural optimisation} \label{sec:nn_qaoa_training}

In this section, we move to a different methodology to find optimal $\QAOA$ parameters than those presented in the previous section. Specifically, as with the incorporation of graph neural networks in the initialisation of the algorithms, we can test neural network based methods for the optimisation itself. Specifically, we test two proposals given in this literature to optimise parameterised quantum circuits. The first, based on a reinforcement learning approach, uses the method of~\cite{khairy_reinforcement-learning-based_2019}. The second is derived from using meta learning to optimise quantum circuits, proposed by~\cite{verdon_learning_2019}. Both of these approaches involve neural networks outputting the optimised parameters by either predicting the update rule or directly predicting the $\QAOA$ parameters.

\subsubsection{Reinforcement learning optimisation} \label{ssec:rl_for_qaoa}
The work of~\cite{khairy_reinforcement-learning-based_2019} frames the $\QAOA$ optimisation as a reinforcement learning problem, adapting~\cite{li_learning_2016} to the problem specific nature of $\QAOA$. The primary idea is to construct and learn a \emph{policy}, $\pi(\boldsymbol{a}, \boldsymbol{s})$, via which an reinforcement learning \emph{agent} associates a \emph{state}, $\boldsymbol{s}_t \in \mathcal{S}$ to an \emph{action}, $\boldsymbol{a}_t \in \mathcal{A}$. In~\cite{khairy_reinforcement-learning-based_2019}, an action is the update applied to the parameters (similarly to~\eqnref{eqn:opt_update_rule}), $\Delta\boldsymbol{\gamma}, \Delta\boldsymbol{\beta}$. A state, $s_t = \mathcal{S}$, consists of the finite differences of the $\QAOA$ cost, $\Delta\mathsf{C}(\paramtheta_{tl})$ and the parameters, $\Delta\boldsymbol{\gamma}_{tl}, \Delta\boldsymbol{\beta}_{tl}$. Here, $l \in \{t-1, \dots, t-L\}$ ranges over the previous $L$ history iterations to the current iteration, $t$. The possible corresponding actions, $\boldsymbol{a}_t \in \mathcal{A}$, are the set of parameter differences, $\{\Delta\boldsymbol{\gamma}_{tl}, \Delta\boldsymbol{\beta}_{tl}\}_{l=t-1}$. The goal of the reinforcement learning agent is to maximise the \emph{reward}, $\mathcal{R}(\boldsymbol{s}_t, \boldsymbol{a}_t, \boldsymbol{s}_{t+1})$ which in this case is the change of $\mathsf{C}$ between two consecutive iterations, $t$ and $t+1$. The agent will aim to maximise a discounted version of the total reward over iterations.

The specific approach used to search for a policy proposed by~\cite{khairy_reinforcement-learning-based_2019} is an actor-critic network in the proximal policy optimisation (PPO) algorithm~\cite{schulman_proximal_2017}, and a fully connected two hidden layer perceptron with 64 neurons for both actor and critic. The authors observed an eight-fold improvement in the approximation ratio compared to the gradient-free Nelder-Mead optimiser\footnote{This was achieved by a hybrid approach where Nelder-Mead was applied to optimise further after a near-optimal set of parameters were found by the RL agent.}. Furthermore, the ability of this method to generalise across different graph sizes is reminiscent of our above $\QAOA$ initialisation approach using GNNs.

\subsubsection{Meta-learning optimisation} \label{ssec:ml_for_qaoa}

A second method to incorporate neural networks is via \emph{meta-learning}. In the classical realm, this is commonly used in the form of one neural network predicting parameters for another. The method we adopt here is one proposed by~\cite{verdon_learning_2019, wilson_optimizing_2021} which uses a neural optimiser that, when given information about the current state of the optimisation routine, proposes a new set of parameters for the quantum algorithm. Specifically,~\cite{verdon_learning_2019, wilson_optimizing_2021} adopts a long short term memory (LSTM) as a neural optimiser (with trainable parameters, $\boldsymbol{\varphi}$), an example of a recurrent neural network (RNN). Using this architecture, the parameters at iteration, $t+1$, are output as:
\begin{equation} \label{eqn:rnn_update_rule}
    \boldsymbol{s}_{t+1}, \paramtheta_{t+1} = \text{RNN}_{\boldsymbol{\varphi}}(\boldsymbol{s}_{t}, \paramtheta_{t}, \mathsf{C}_t)
\end{equation}

\begin{figure}[!htb]
\centering
\begin{tikzpicture}
\begin{axis}[
    xlabel={Epochs},
    ylabel={Approximation ratio, $r$},
    xmin=0, xmax=130,
    ymin=0, ymax= 1,
    xtick={0, 25, 50, 75, 100, 125}, 
    ytick={0, 0.1, 0.2, 0.3, 0.4, 0.5, 0.6, 0.7, 0.8, 0.9, 1.0},
    legend pos=south east,
    ymajorgrids=true,
    legend cell align={left},
    grid =major,
    height= 6cm,
    width=7.5cm,
]
\addplot[
    color=red,
    mark=square*,
    ]
    coordinates {
(0,  0.03 )   
(25, 0.28)
(50,  0.45 )
(70, 0.72)
(85,	0.85)
(100,    0.88)
(125,	 0.91)
(150,	0.92)
(175,	0.93)

};

\addplot[
    color=green,
    mark=triangle*,
    ]
    coordinates {
(0,  0.03 )    
(20,  0.87 )
(25,  0.96 )
(30,  0.92 )
(35,  0.94)
(40,  0.95)
(45,  0.95)
(50,  0.95)
(55,  0.96)
(60,  0.96)
(65,  0.96)
(70,  0.96)

};

\addplot[
    color=blue,
    mark=square*,
    ]
    coordinates {
(0,  0.03 )    
(35,  0.83 )
(40,  0.90 )
(45,	0.93)
(50,	0.88)
(55,	0.91)
(60,	0.93)
(70,	0.93)
(75,	 0.94)
(80,	0.94)
(95,	0.94)
(110,	0.94)
(125,	 0.94)

};

\legend{SGD, Meta-learning + SGD,  RL + SGD}
\end{axis}
\end{tikzpicture}
\caption{\textbf{Comparison of neural optimisers}. We compare the LSTM based meta-learning against a reinforcement learning optimiser, with vanilla stochastic gradient descent (SGD) used as a benchmark. Once each neural optimiser has converged, we continue the optimisation with SGD. Here we use $10$ qubits for $\maxcut$ and $\QAOA$ depth $p=6$. The results are also averaged over independent $10$ runs.}
\label{fig:neural_optimisers}
\end{figure}
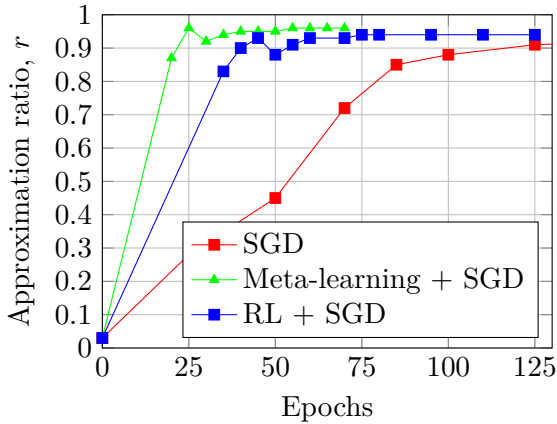

Here, $\boldsymbol{s}_{t}$ is the hidden state of the LSTM at iteration $t$, and the next state is also suggested by the neural optimiser with the $\QAOA$ parameters. $\mathsf{C}_t$ is used as a training input to the neural optimiser, which in the case of a VQA is an approximation to the expectation value of the problem Hamiltonian, i.e.~\eqnref{eqn:maxcut_hamiltonian_quantum}. The cost function for the RNN chosen by~\cite{verdon_learning_2019} incorporates the averaged history of the cost at previous iterations as well as a term that encourages exploration of the parameter landscape. We compare this approach against SGD (with a fixed learning rate of $\eta = 0.01$) in~\figref{fig:neural_optimisers} and the previous RL based approach.

\section{Conclusion and Outlook} \label{sec:conclusion}
The work presented in this paper builds new techniques into analysing the $\QAOA$ algorithm - a quantum algorithm for constrained optimisation problems. Here, we build an efficient and differentiable process using the powerful machinery of graph neural networks. GNNs have been extensively studied in the classical domain for a variety of graph problems and we adopt them as an initialisation technique for the $\QAOA$, a necessary step to ensure the $\QAOA$ is capable of finding solutions efficiently. Good initialisation techniques are especially crucial for variational algorithms to achieve good performance when implemented on depth-limited near term quantum hardware. Contrary to the previous works on $\QAOA$ initialisation, our GNN approach does not require separate instances each time one encounters a new problem graph and therefore can speed up inference time across graphs. We demonstrated this in the case of the $\QAOA$ by showing good generalisation capabilities on new (even larger) test graphs than the family of graphs they have been trained on. To complement the initialisation of the algorithm, we investigate the search for optimal $\QAOA$ parameters, or optimisation, with a variety of methods. In particular, we incorporate gradient-based/gradient-free classical and quantum-aware optimisers along with more sophisticated optimisation methods incorporating meta- and reinforcement learning.

There is a large scope for future work, particularly in the further incorporation and investigation of classical machine learning techniques and models to improve near term quantum algorithms. One could consider alternative GNN structures for initialisation, other neural optimisers or utilising transfer learning techniques. For example, one could study the combination of warm-starting abilities of GNNs with generic quantum circuit initialisers such as FLIP~\cite{sauvage_flip_2021}, both of which exhibit generalisation capabilities to larger problem sizes.

A second extension of our proposal could be the extension to other graph problems besides simply $\maxcut$, and at much larger scales following~\cite{schuetz_combinatorial_2022}. Finally, one could consider the use of truly `quantum' machine learning models such as quantum graph neural networks~\cite{verdon_quantum_2019-2} or others~\cite{benedetti_parameterized_2019, abbas_power_2021}.

\section*{Acknowledgements}

The results of this paper were produced using Tensorflow Quantum~\cite{broughton_tensorflow_2020} and the corresponding code can be found at~\url{https://github.com/nishant34/GNNS-for-initializing-QAOAs.git}. NJ wrote the code and ran the experiments. BC, EK and NK supervised the project. All authors contributed to manuscript writing. We thank Stefan Sack for helpful correspondence. We acknowledge funding from EPRSC grant EP/T001062/1. For the purpose of open access, the author has applied a Creative Commons Attribution (CC BY) licence to any Author Accepted Manuscript version arising from this submission.



\newpage
\onecolumngrid

\appendix

\section{Graph neural network architectures} \label{app:gnn_architectures}

\subsection{Line graph neural network}\label{app:lgnn}
As mentioned in the main text, the primary GNN architecture we choose is a \emph{line} graph neural network, also adopted by~\cite{yao_experimental_2019} for combinatorial optimisation and proposed by~\cite{chen_supervised_2019}. Given a graph, $\mathcal{G} := \left(\mathcal{V}_{\G}, \mathcal{E}_{\G}\right)$ with vertices $\mathcal{V}_{\G}$ and edges $\mathcal{E}_{\G} = \{(i, j) | i, j \in \mathcal{V}_{\G} \text{ and } i \neq j\}$, the line graph, denoted $\LG$, is then constructed by taking the edges of $\mathcal{G}$ which become the nodes of $\LG$, $\mathcal{E}_{\G} \rightarrow \mathcal{V}_{\LG}$. $\LG$ only has an edge between two vertices if the vertices shared a node in the original graph, $\mathcal{G}$. For example, if we had three nodes in $\G$, $a, b$ and $c$ which were connected as $a - b$ and $b - c$, the vertex set of $\LG$ would contain nodes labelled $(a-b)$ and $(b-c)$ and would have a edge between them since they both contain the vertex $b$. This behaviour is described by a `non-backtracking' operator~\cite{chen_supervised_2019, yao_experimental_2019} introduced by~\cite{krzakala_spectral_2013} and enables information to propagate in a directed fashion on $\LG$.

The LGNN then actually contains two separate graph neural networks, one defined on $\G$ and another defined on $\LG$. The GNN on $\G$ has feature vectors, $\hbs^{n_{\nu}}_{t} \in \mathbb{R}^d$ for each node $n_{\nu}$ and each iteration, $t$. Similarly the GNN on $\LG$ has feature vectors, $\gbs^{n_{\mu}}_{t}$ for every node $n_{\mu}$ in $\LG$.

Without the information from the line graph, the feature vectors for $\G$ would be updated as:
\begin{align}         
    \Bar{\ybs}^{n_{\nu}}_{t+1} &:= \hbs^{n_{\nu}}_{t}\paramtheta^0_t + \mathcal{D}\hbs^{n_{\nu}}_{t}\paramtheta^1_t + \sum_{j=1}^J\mathcal{A}_j\hbs^{n_{\nu}}_{t}\paramtheta^j_t\\
    \ybs^{n_{\nu}}_{t+1} & := f\left(\Bar{\ybs}^{n_{\nu}}_{t+1}\right)\\
    \hbs^{n_{\nu}}_{t+1} &= [\ybs^{n_{\nu}}_{t+1},\Bar{\ybs}^{n_{\nu}}_{t+1}]
\end{align}
Where $\hbs^{n_{\nu}}_{t+1}$ results from the concatenation of the two vectors, $\ybs^{n_{\nu}}_{t+1},\Bar{\ybs}^{n_{\nu}}_{t+1}$. $\mathcal{D}$ is the degree matrix of $\mathcal{G}$ and $[\mathcal{A}_j]_{lm} := \min(1, [\mathcal{A}^{2^j}]_{lm})$ are power graph adjacency matrices (where $\mathcal{A}$ is the adjacency matrix of $\G$), which allows information to be aggregated from different neighbourhoods. The matrix element $[\mathcal{A}^{2^j}]_{lm}$ gives the number of walks between node $l$ and node $m$ of length $2^j$ and $\mathcal{A}_j$ converts this information into a binary matrix describing whether a walk \emph{exists} (of length $2^j$) between $l$ and $m$. $f$ is a nonlinear function, taken in~\cite{chen_supervised_2019, yao_experimental_2019} to be ReLu, $f(x) = \max(0, x)$

Now, including updates from the line graph into the GNN, the feature vectors from each graph are updated in tandem as follows:
\begin{align}         
    \Bar{\ybs}^{n_{\nu}}_{t+1} &:= \hbs^{n_{\nu}}_{t}\paramtheta^0_t + \mathcal{D}\hbs^{n_{\nu}}_{t}\paramtheta^1_t + \sum_{j=1}^J\mathcal{A}_j\hbs^{n_{\nu}}_{t}\paramtheta^j_t + \mathcal{S}\gbs^{n_{\mu}}_{t}\paramtheta^{J+1}_t + \mathcal{U}\gbs^{n_{\mu}}_{t}\paramtheta^{J+2}_t \\
    \Bar{\zbs}^{n_{\mu}}_{t+1} &:= \gbs^{n_{\mu}}_{t}\boldsymbol{\varphi}^0_t + \mathcal{D}_{\LG}\gbs^{n_{\mu}}_{t}\boldsymbol{\varphi}^1_t + \sum_{j=1}^J\mathcal{B}_j\gbs^{n_{\mu}}_{t}\boldsymbol{\varphi}^j_t + \mathcal{S}\hbs^{n_{\nu}}_{t}\boldsymbol{\varphi}^{J+1}_t + \mathcal{U}\hbs^{n_{\nu}}_{t}\boldsymbol{\varphi}^{J+2}_t \\
    \ybs^{n_{\nu}}_{t+1} & := f\left(\Bar{\ybs}^{n_{\nu}}_{t+1}\right), \qquad  \zbs^{n_{\mu}}_{t+1}  := f\left(\Bar{\zbs}^{n_{\mu}}_{t+1}\right)\\
    \hbs^{n_{\nu}}_{t+1} &= [\ybs^{n_{\nu}}_{t+1},\Bar{\ybs}^{n_{\nu}}_{t+1}], \qquad \gbs^{n_{\mu}}_{t+1} = [\zbs^{n_{\mu}}_{t+1},\Bar{\zbs}^{n_{\mu}}_{t+1}]
\end{align}
Here, $\paramtheta, \boldsymbol{\varphi}$ are the trainable parameters of the GNN over $\mathcal{G}$ and the GNN over $\LG$ respectively. The matrices, $\mathcal{S}, \mathcal{U}$ are signed and unsigned \emph{incidence matrices}. These are defined for every node $i$ of $\mathcal{G}$ and the nodes $(k \rightarrow l)$ of $\LG$ (which are the edges of $\G$) as:
\begin{equation}
    \mathcal{U}_{i, (k\rightarrow l)} = 
    \begin{cases}
        1 \text{ if }  i = k\\
        0 \text{ otherwise }
    \end{cases}, \qquad    \mathcal{S}_{i, (k\rightarrow l)} = 
    \begin{cases}
        1 \text{ if }  i = k\\
        -1 \text{ if } i = l\\
        0 \text{ otherwise }
    \end{cases}
\end{equation}
Finally, $\mathcal{B}$ is the non-backtracking operator describing $\LG$, defined as:
\begin{equation}
    \mathcal{B}_{(i\rightarrow j), (k\rightarrow l)} = 
    \begin{cases}
        1 \text{ if } j = k \text{ and } i \neq l\\
        0 \text{ otherwise }
    \end{cases}
\end{equation}
with its power graphs, $\mathcal{B}_{j}$, defined analogously to $\mathcal{A}_{j}$.

\subsection{Graph convolutional network}\label{app:gcn}
The alternative architecture chosen by~\cite{schuetz_combinatorial_2022} is the graph convolutional network architecture, which is simpler than the line graph neural network above.
Here, the embedding vector updates have the following form:
\begin{equation}\label{eqn:gcnn_equation}
    \hbs^{n_{\nu}}_{t+1} = f\left(\paramtheta^0_t\sum_{j: n_{j}\in \mathcal{N}(n_{\nu})}\frac{\hbs^{n_{j}}_{t}}{|\mathcal{N}(n_{\nu})|} + \paramtheta^1_t\hbs^{n_{\nu}}_{t}\right)
\end{equation}
where $\mathcal{N}(n_{\nu})$ is the local neighbourhood of $n_{\nu}$, $\mathcal{N}(n_{\nu}) = \{n_j \in \mathcal{V}|(n_{\nu}, n_{j}) \in \mathcal{E}\}$ and $|\cdot|$ denotes cardinality. This is a simpler architecture since the updates in a single step only involve information passing from the immediately local nodes to a given one, whereas the LGNN involves contributions across the graph in a single step.

\end{document}